\documentclass[aps,amsmath,amssymb,pre,showpacs]{revtex4}

\usepackage{graphicx}
\usepackage{epsfig}

\newcommand{\be}{\begin{equation}}
\newcommand{\ee}{\end{equation}}
\newcommand{\bea}{\begin{eqnarray}}
\newcommand{\eea}{\end{eqnarray}}
\newcommand{\bay}{\begin{array}}
\newcommand{\eay}{\end{array}}
\newcommand{\bfig}{\begin{figure}}
\newcommand{\efig}{\end{figure}}
\newcommand{\bc}{\begin{center}}
\newcommand{\ec}{\end{center}}
\newcommand{\btab}{\begin{tabular}}
\newcommand{\etab}{\end{tabular}}

\newcommand{\dr}{\partial}

\newcommand{\al}{\alpha}
\newcommand{\bt}{\beta}

\begin{document}

\preprint{draft}

\title{Dynamics of granular avalanches caused by local perturbations}

\author{Thorsten Emig}
\affiliation{
Institut f\"ur Theoretische Physik, Universit\"at zu K\"oln,\\
Z\"ulpicher Stra\ss e 77, D-50937 K\"oln, Germany}

\author{Philippe Claudin}
\affiliation{
Laboratoire de Physique et M\'ecanique des Milieux H\'et\'erog\`enes,\\
ESPCI, 10 rue Vauquelin, 75231 Paris Cedex 05, France.}

\author{Jean-Philippe Bouchaud} 
\affiliation{
Service de Physique de l'{\'E}tat Condens{\'e}, Centre d'{\'e}tudes de Saclay\\
Orme des Merisiers, 91191 Gif-sur-Yvette Cedex, France}

\date{\today}

\begin{abstract}
  Surface flow of granular material is investigated within a continuum
  approach in two dimensions. The dynamics is described by a
  non-linear coupling between the two `states' of the granular
  material: a mobile layer and a static bed. Following previous
  studies, we use mass and momentum conservation to derive St-Venant
  like equations for the evolution of the thickness $R$ of the mobile
  layer and the profile $Z$ of the static bed. This approach allows
  the rheology in the flowing layer to be specified independently, and
  we consider in details the two following models: a constant plug
  flow and a linear velocity profile.  We study and compare these
  models for non-stationary avalanches triggered by a localized amount
  of mobile grains on a static bed of constant slope. We solve
  analytically the non-linear dynamical equations by the method of
  characteristics. This enables us to investigate the temporal
  evolution of the avalanche size, amplitude and shape as a function
  of model parameters and initial conditions. In particular, we can
  compute their large time behavior as well as the condition for the
  formation of shocks.
\end{abstract}

\pacs{
45.70.Ht --- Avalanches,
45.70.-n --- Granular systems,
05.45.-a --- Nonlinear dynamics and nonlinear dynamical systems
}

\maketitle

\section{Introduction}
\label{sec:intro}

The dynamics of granular avalanches has been keeping busy a large
fraction of the granular community for several years. As an example, a
recent review paper has been written by the French research group
named `GdR Milieux Divis\'es' in order to sum up the `French results'
of experiments and numerical simulations on steady uniform dense
granular flows \cite{GdR}. Some of the recurring issues addressed in
this long article concern the definition and the description of the
rheology and the friction law of these flows. As a matter of fact,
these quantities are important physical ingredients to be plugged into
the equations proposed for the modeling of these avalanches.
Although some alternative models are proposed -- see e.g.
\cite{AT01,R02} -- an interesting theoretical framework for the
description of granular flows follows the St Venant-like approach for
thin flows in hydrodynamics, in which conservation equations are
integrated over the depth of the flow \cite{LL-vol6}. This was
proposed for example by Savage, Hutter and co-workers for the case of
the motion of grains over an inclined rough plane \cite{SH89}.

With such equations adjusted on uniform steady flows, it is for
instance possible to reproduce quantitatively steady fronts of
granular avalanches on a rough inclined plane \cite{P99}. Another
interesting situation is that of unsteady avalanches, whose dynamics
is of course more complicated to capture and is very demanding for
the models. An example of such a situation is the motion of an
initially confined granular mass which is released and runs down the
same rough plane. In this case, the same calibration gives also quite
good qualitative predictions \cite{PF02}.  The spreading of a granular
mass is also of interest in a geophysical context as they are a model
of real cliff collapse \cite{LMV04,MBLAVP04}. Another famous example
is the case where an initial static layer of grains is available on
the plane, and avalanches triggered by a local solicitation with a
pointy object \cite{DD99,D01}. The nice experimental pictures show
two types of behavior: a `triangular avalanche' when the thickness of
the grain layer is small, and an avalanche with an `up-hill'
propagative front when the layer is thicker.

%
\begin{figure}[t]
\begin{center}
\includegraphics[height=1.5in]{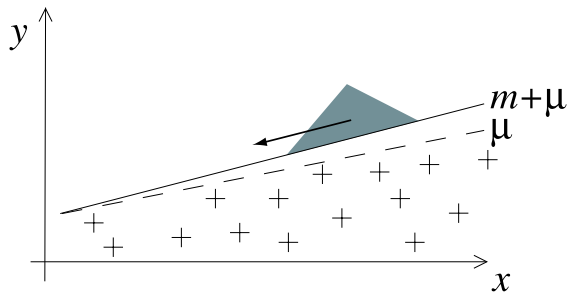}
\caption{Rolling grains are locally deposited on a static sand bed which has 
  initially a uniform slope $\mu+m$, where $\mu$ is the tangent of the
  angle of repose.
\label{fig:setup}}
\end{center}
\end{figure}
%

In the present paper, our aim is also to deal with unsteady and
non-uniform granular flows in a geometry close to these elementary
`response' situations where avalanches are caused by local
perturbations. More precisely, we look at the dynamical evolution of a
uniform slope on which some amount of rolling grains is initially
deposited locally. However, unlike the above examples, we shall focus
on avalanches running over an erodible bed, i.e. an infinite grain
layer for which the selected thickness of moving grains is not fixed
\textit{a priori} or bounded by a plane. This situation then has one
extra dynamical degree of freedom.  As will be made more precise in the
next section, in order to close these equations it is indeed necessary
to specify the shape of the velocity profile of the moving grains and
to explicit the different forces acting on them.  Moreover, as
our theoretical work aims at obtaining exact results on avalanche profiles, 
the analysis will be restricted to
two dimensional systems for simplicity, and will be relevant for
experiments in confined cells or when the perturbation has translation
invariance in the direction perpendicular to the slope.

General St Venant equations for granular avalanches on erodible beds
have been recently proposed by Douady \textit{et al.} in \cite{DAD99}.
Observations show that these avalanches consist of a thin moving layer
of grains over a (quasi)static profile \cite{KINN01}. The two natural
variables in this modeling are therefore the profile of the static
pile $Z$, and the thickness of rolling grain layer $R$ that flow over
this bed, both functions of time and horizontal position. In this
framework, we shall derive two particular sets of coupled equations
for $R$ and $Z$ that are simple enough to be solved exactly for the
`response geometry' described above. As will be explicited in the next
section, they are based on different choices for the velocity profile
in the flowing grain layer, and exhibit different non-linearities. A
particular case of one of the two studied sets gives in fact the
equations proposed in \cite{BCRE94} on more phenomenological grounds,
and for which we developed an analytical treatment in \cite{ECB00}
using the method of characteristics. We extend here this powerful
technique to all other cases as well. This technique has been also
applied to a St Venant like model for debris avalanches \cite{MHR00}.

After the derivation of these sets of equations in section \ref{sec:StVenant}
in which we shall specify the assumptions made, both sets of equations 
-- corresponding to different flow profiles -- are treated
in the following two sections. We show the different predictions of the models,
compare how the triggered avalanche dies out or grows depending whether the
initial slope of the static profile is smaller or larger than the repose
angle of the pile, and discuss the appearance of shocks. We finally
conclude and draw perspectives on the description of avalanche fronts.
The present paper is rather technical, but we hope that the scope of the method used is 
sufficiently broad to warrant interest in itself; furthermore, obtaining explicit exact
result for the shape of avalanche shapes and sizes gives important benchmarks to which
experiments can be compared, and will eventually help selecting the correct set of hydrodynamical equations 
for granular avalanches.

\section{St Venant equations for granular avalanches}
\label{sec:StVenant}

\begin{figure}[t]
\begin{center}
\includegraphics[height=2.5in]{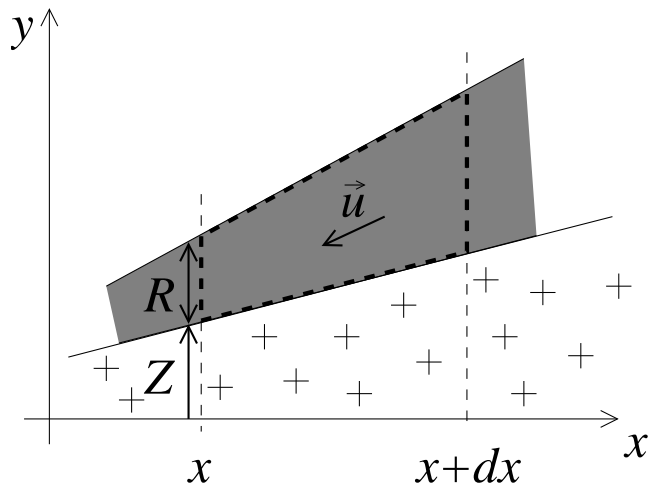}
\caption{Rolling phase $R(t,x)$ (grey) and static phase $Z(t,x)$ (crosses). 
  The grains flow from the right to the left. The dashed box is the
  control volume for which mass and momentum conservation laws are written.
\label{schema}}
\end{center}
\end{figure}

The aim of this section is to (re-)establish coupled differential equations
for the variables $R$ and $Z$ introduced above and depicted in figure
\ref{schema}. These equations  encode the conservation of the
number of grains as well as their horizontal momentum \cite{DAD99}. They describe
how static grains may be dislodged and contribute to the flow
(erosion), and vice-versa how moving grains may come to rest
(deposition) \cite{BCRE94}. 

Besides $R$ and $Z$, another important field is of course the velocity
in the moving layer.  Let us note $u(x,y)$ the horizontal component of
the velocity profile in this layer -- although not explicited, time
dependence of $u$ is understood.  The averaged velocity of
the flow can be defined as $U=\frac{1}{R} \int_Z^{Z+R} \! dy \,
u(x,y)$. With this quantity, the number of particles passing through
a vertical line in $x$ during the time interval $dt$ is
$\frac{\rho}{m} RU dt$, where $\rho$ is the mass density of the
granular material and $m$ the mass of a single grain. The difference
of this number in $x$ and $x+dx$ makes the volume $(R+Z)dx$ change, so
that the conservation of matter reads
\begin{equation}
\label{conservmatter}
\dr_t(R+Z) = \dr_x(RU).  
\end{equation} 
In a similar way, the $x$-component of the momentum passing through the vertical
line in $x$ during $dt$ is $\rho RW dt$, where $W$ is the average of the
square of the velocity: $W=\frac{1}{R} \int_Z^{Z+R} \! dy \, u^2(x,y)$.
The change of horizontal momentum $\rho RU dx$ in the control volume
(see figure \ref{schema}) can then be written as
\begin{equation}
\label{conservmomentum1}
\dr_t (RU) \, dx = \dr_x (RW) \, dx + \frac{1}{\rho} \, dF_x,
\end{equation}
where $dF_x$ is the $x-$component of the forces acting of this
volume.

%
%

These forces are of two kinds: those coming from the lateral sides
through the stress $\sigma_{xx}$, and those due to the contact between
the rolling and the static phase. Assuming a horizontal normal stress
proportional to the `hydrostatic' pressure $\sigma_{xx}(x,y)=b \, \rho
g (R+Z-y)$, the corresponding force is $\int_Z^{Z+R} \! dy \,
\sigma_{xx} = \frac{1}{2} b \rho g R^2$. Numerical simulations suggest
values for $b$ that increase from roughly one half in static piling to
unity for uniform steady flows \cite{GdR,dC04,SEGHLP01}.  In
\cite{SH89} the value of $b$ is obtain for unsteady and inhomogeneous
situations from a Mohr-Coulomb yield criterion. It turns out that $b$
can increase (decrease) compared to unity if the grains are compressed
(decompressed) by the flow.  Here we consider $b$ as a
phenomenological {\it constant}.  Taking the difference of this force
on the two sides of the control volume, the contribution of the
horizontal normal stress to $dF_x$ finally reads $b \rho g R \dr_x R
\, dx$. To compute the other part of $dF_x$, we assume that the grains
in the control volume behave, with respect to the static phase, like a
frictional solid (with normal reaction $N$ and friction force $T$). We
call $\mu$ the friction coefficient.  Defining $\theta$ by
$\tan\theta=\dr_x Z$, the balance of the weight of this volume gives
$\rho g R \, dx = N \cos\theta + T \sin\theta$. With $T=\mu N$, the
horizontal contribution of $N$ and $T$ to $dF_x$ (to the lowest order
in $\theta$ and $\mu$) is $\rho g R (\dr_x Z - \mu) \, dx$.  In
summary, we get
\begin{equation}
\label{dF_x}
dF_x = b \rho g R \dr_x R \, dx + \rho g R (\dr_x Z - \mu) \, dx.
\end{equation}

Several remarks are in order at this point. First of all, we have
assumed that the granular density $\rho$ is the same in the moving and
the static parts, which is certainly a valid hypothesis for dense
flows as considered here. More importantly, the friction coefficient
$\mu$ will be taken as a constant in the following.  It is however
well established that there are few degrees of hysteresis between the
starting and the stopping angles, which can be simply understood even
at the scale of a single grain rolling down a ``pile'' consisting of a
layer of regularly spaced fixed grains \cite{QADD00}. As will be
emphasized in the conclusion, this slight difference is of great
importance when it comes to the description of regions where the
moving layer is about to jam or the static bed about to move, i.e. at
the feet of avalanche fronts ($R \to 0$).  Besides, it has been shown
that in quasi 2D experiments conducted in a thin channel between two
plates, the friction of the rolling on those plates plays a major role
in the dynamics of the avalanche \cite{BDL02,TRVLPJD03}.  Such a
boundary effect is not encoded here.

Finally, these equations will be closed by making an explicit choice
for the velocity profile $u$. The integrals for the computation of $U$
and $W$ will then be expressed as a function of $R$. This means that
we implicitly assume {\it instantaneous} adaptation of the velocity to
the flow. In principle, like $R$ and $Z$, the averaged velocity $U$ is
an independent dynamical variable whose dynamics should be described
by an additional equation of `internal force balance' $\dr_t U =
\ldots$.  The right hand side would specify the part of the forces
acting on the moving layer that contribute to the acceleration and
friction of the rolling grains rather than the erosion/deposition
processes, i.e., the exchange between $R$ and $Z$. A model similar to
our Eqs.~(\ref{conservmatter}) - (\ref{dF_x}) has been very recently
studied by Khakhar et al. \cite{KOO04} but for different geometries
and initial conditions (heap and rotating cylinder flows).  Their
momentum balance equation has however an additional term which
describes grain collisions in the mobile phase. Thus they combine
effects of internal dissipation \cite{KMSO97} and external forces as
gravity and solid like friction between the two phases. 

Two choices for the velocity profile will be investigated below: the
plug flow for which $u(x,y)=U=\mbox{Cst}$ (named model $\mathcal{P}$
for `plug'), and the situation with a linear velocity profile
$u(x,y)=\gamma y$ (named model $\mathcal{L}$ for `linear'). The
natural selection of this profile is still a puzzling issue and is the
result of several mechanisms at the grain level (e.g. trapping) or at
larger scales (non-local effects, clustering) \cite{AD01,BDLBB02}. At
the phenomenological level, the first case is reasonable for the
description of thin or dilute flows (when $R$ of the order of a grain diameter)
\cite{ARdG99}, and the second one is strongly supported by experiments
and simulations performed on \emph{steady and deep} 2D systems
\cite{GdR}.  In particular, it should be emphasized that the velocity
gradient is found to be constant, and it is the thickness of the
rolling layer $R$ which adapts its value to the external imposed shear
stress, see e.g.  \cite{GdR,KOAO01}.  In contrast, note that for
steady granular flows on a fixed rough inclined plane a Bagnold-like
velocity profile $U \propto R^{3/2}$ is observed. For the study of
unsteady situations, since the static profile $Z$ is now specified, the two St Venant
conservation equations can be used for the determination of the space and
time evolution of $R$ and $U$ \cite{PF02}.

\section{Plug flow: constant velocity profile (Model $\mathcal{P}$)}
\label{sec:constant_profile}

For a plug flow with a constant average horizontal velocity $U$ we
trivially get $W=U^2$, and one obtains from the analysis in
section \ref{sec:StVenant} the model
\begin{subequations}
\label{eq:SVUconstant}
\begin{eqnarray}
\label{SVUconstfinalZ}
\dr_t Z & = & - R \, \dr_x Z - b \, R \, \dr_x R,\\
\label{SVUconstfinalR}
\dr_t R & = & \dr_x R + R \, \dr_x Z + b \, R \, \dr_x R.
\end{eqnarray}
\end{subequations}
Here the dimensionless $t$, $x$, $R$ and $Z$ are measured in units of
$U^2/g$, and $t$ is rescaled by $U/g$. Furthermore, as we are
interested in surface profiles close to the avalanche slope $\mu$, $Z$
is measured relative to the critical slope, i.e., it is replaced by $Z
+ \mu x$. The quantity $b$ is then the only free parameter of these
equations.  Recall that an isotropic stress distribution corresponds
to $b=1$.  Note also that setting $b=0$ yields the so-called BCRE
model introduced in \cite{BCRE94} (although without the diffusive
terms considered there). Any small but finite $b$ thus leads to novel
non-linearities.  Before we study the propagation of a localized
perturbation, as a first simple benchmark of the model, we briefly
note the predictions of the model for the initial situation of a
constant slope $Z_0(x)=mx$ and a homogeneous amount of rolling grains,
$R_0(x)=\varrho$. It is easy to see that in this case the thickness of
the mobile layer growths (or decays) exponentially in time, depending
on the sign of $m$, with $R(t,x)=\varrho e^{mt}$.  The static
profile is then given by $Z(t,x)=mx+\varrho(1-e^{mt})$.  Note that
this solution does {\it not} depend on $b$ since $R(t,x)$ is
independent of $x$.

\subsection{Infinite stress anisotropy ($b=0$)}
\label{sec:constant_b_zero}

\subsubsection{Analytical solution} 

\begin{figure}[t]
\begin{center}
\includegraphics[height=2.5in]{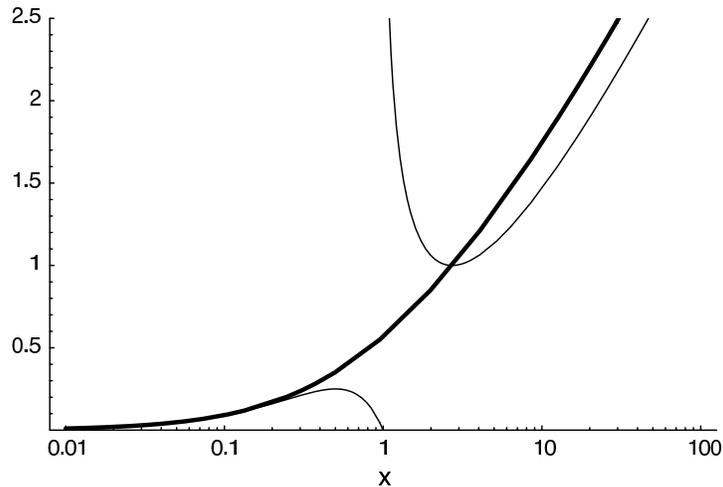}
\caption{Bold curve: plot of the Lambert's function $W(x)$. Thin curves: 
small and large $x$ expansions $W(x) \sim x - x^2$ and 
$W(x) \sim \ln x - \ln \ln x$, respectively. 
\label{fig:Lambert}}
\end{center}
\end{figure}

This non-linear model has been studied previously for the case of a
static profile consisting of two regions of constant but different
slopes and an initially homogeneous (constant) amount of rolling
grains \cite{ECB00}.  However, the method of characteristic curves can
be also employed to study local perturbations in form of an initially
localized amount $R_0(x)$ of rolling grains. In fact, for arbitrary
initial profiles $R_0(x)$ and $Z_0(x)$ the solution of equations
(\ref{eq:SVUconstant}) with $b=0$ can be obtained analytically in
implicit form. Using the method of characteristic curves, see Appendix
\ref{sec:app_cc}, we introduce the new coordinates $\alpha(t,x)$ and
$\beta(t,x)$. The mapping between the two coordinate systems depends
on the solution for $R(t,x)$, $Z(t,x)$ and thus on the initial
profiles due to the non-linearities.  The so-called characteristic
equations, which determine simultaneously the coordinate mapping and
the profiles, for this model read
\begin{subequations}
\label{eq:bcre_b_0_ceqs}
\begin{eqnarray}
\label{eq:bcre_b_0_characcoord+}
\dr_\al x - \zeta_+ \dr_\al t & = & 0,\\
\label{eq:bcre_b_0_characcoord-}
\dr_\bt x - \zeta_- \dr_\bt t & = & 0,\\
\label{eq:bcre_b_0_characSV+}
- R \dr_\al Z + (\zeta_+ - R) \, \dr_\al R & = & 0,\\
\label{eq:bcre_b_0_characSV-}
- R \dr_\bt Z + (\zeta_- - R) \, \dr_\bt R & = & 0,
\end{eqnarray}
\end{subequations}
with the characteristic directions given by
\begin{equation}
\label{eq:bcre_b_0_zeta}
\zeta_+=-1, \quad \zeta_-=R.
\end{equation}
The solution of these equations can be obtained explicitly, 
\begin{subequations}
\label{eq:BCRE-map}
\begin{eqnarray}
  \label{eq:BCRE-map:t}
  t(\alpha,\beta) &=& -\int_{-\alpha}^\beta
  \frac{d\beta'}{1+R(\alpha,\beta')} \\
  \label{eq:BCRE-map:x}
  x(\alpha,\beta) &=& -\beta - t(\alpha,\beta). 
\end{eqnarray}
\end{subequations}
As function of these new coordinates the solution can be written in
the closed form
\begin{subequations}
\label{eq:BCRE-b=0-solution}
\begin{eqnarray}
  \label{eq:BCRE-b=0-solution-Z}
  Z(\alpha,\beta) &=& Z_0(\alpha) \\
  \label{eq:BCRE-b=0-solution-R}
  R(\alpha,\beta) &=& W\left[R_0(-\beta) e^{
R_0(-\beta) + Z_0(-\beta) - Z_0(\alpha)} \right],
\end{eqnarray}
\end{subequations}
where $W$ is Lambert's function \cite{Lambert}, see
Fig.~\ref{fig:Lambert}.  Before we proceed with special choices for
localized initial profiles $R_0(x)$, we would like to point out that
the model of equations (\ref{eq:SVUconstant}) with $b=0$ can be still
solved exactly if the coefficient of $\partial_x Z$ on the right hand
side of both equations is a general function of $R(t,x)$, ${\mathcal
  F}(R)$. The above case corresponds to ${\mathcal F}(R)=R$.

%
%

\begin{figure}[h] 
\includegraphics[width=0.4\linewidth]{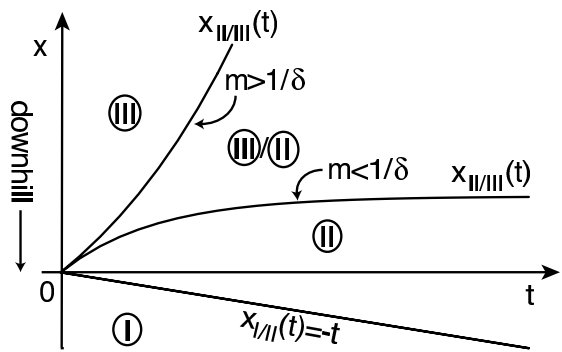} 
\caption{Model $\mathcal{P}$: Boundaries between different sectors 
in the $t-x$ plane.}
\label{fig:sectors}
\end{figure}

In order to progress with analytical techniques we make a special
choice for a localized $R_0(x)$ which allows for a closed expression
for the integral of the coordinate map in equation
(\ref{eq:BCRE-map:t}).  Below, we will demonstrate by an explicit
numerical integration of the coordinate map for a generic Gaussian
perturbation $R_0(x)$ that the following results are robust with
respect to the precise form of the perturbation. Thus we proceed with
the choice
\begin{equation}
  \label{eq:BCRE-R0}
  R_0(x)=W\left[r_0 e^{r_0- |x|/\delta}
  \right].
\end{equation}
for the initial profile.  This profile decays exponentially at large
$|x|$ and has a amplitude of $r_0$. For the relevant limit of $r_0 \ll
1$ the width at half amplitude is $\delta \ln 2$.  This form is
adapted to the integral in equation (\ref{eq:BCRE-map:t}) since the
latter equation can be written as
\begin{equation}
\label{eq:bcre-b=0-map-t-special}
  t(\alpha,\beta) = -\alpha-\beta + \int_{-\alpha}^\beta 
   \frac{\partial_{\beta'} R(\alpha,\beta') d\beta'}{\partial_{\beta'} 
     (\ln R_0(-\beta')) - R'_0(-\beta') -
     Z'_0(-\beta')}
\end{equation}
and $R_0(x)$ of equation (\ref{eq:BCRE-R0}) has the property that
\begin{equation}
  \label{eq:BCRE-R0-prop}
  \partial_x(\ln R_0(-x)) - R'_0(-x)=-\frac{1}{\delta} 
\mbox{\rm sgn}(x). 
\end{equation}
For the static phase we will consider always a profile with a constant
slope, 
\begin{equation}
  \label{eq:BCRE-Z0}
  Z_0(x)=mx,
\end{equation}
where due to our definition of $Z$ the parameter $m$ measures the {\em
  excess} slope relative to the critical angle. For these initial data
the solution of equations (\ref{eq:SVUconstant}) in the curved coordinate
frame reads
\begin{subequations}
\label{eq:BCRE-R+z}
\begin{eqnarray}
  \label{eq:R-of-ab}
  R(\alpha,\beta)&=& W\left[ r_0 e^{r_0 -
        |\beta|/\delta - m (\alpha+\beta)}\right]\\
  \label{eq:Z-of-ab}
  Z(\alpha,\beta)&=&m\alpha.
\end{eqnarray}
\end{subequations}
In order to integrate the equations for the coordinate mapping, see
equation (\ref{eq:BCRE-map:t}), we have to divide the space-time into
different sectors due to the sign function of
equation (\ref{eq:BCRE-R0-prop}). Depending on the sign of $\alpha$ and
$\beta$ we define the following three sectors, (I): $\alpha \le 0$,
$\beta>0$, (II): $\alpha\le 0$, $\beta\le 0$, and (III): $\alpha >0$,
$\beta\le 0$. The case $\alpha$, $\beta>0$ is mapped to negative
times, and is therefore not of interest. The boundary between the
sectors (II) and (III) in the $t-x$ plane can be obtained by
integrating equation (\ref{eq:BCRE-map:t}) with $\alpha=0$ and $\beta<0$.
The boundaries are given by
\begin{subequations}
\label{eq:BCRE-boundaries}
\begin{eqnarray}
  \label{eq:bound-I-II}
  x_{\rm I/II}(t)&=&-t,\\
  \label{eq:bound-II-III}
 x_{\rm II/III}(t)&=&\frac{r_0}{m-1/\delta}\left(
 e^{(m - 1/\delta)t}-1\right),
\end{eqnarray}
\end{subequations}
where the subscript indicates the adjacent sectors, see
figure \ref{fig:sectors}.

%
%

\begin{figure}[h] 
  \includegraphics[width=0.48\linewidth]{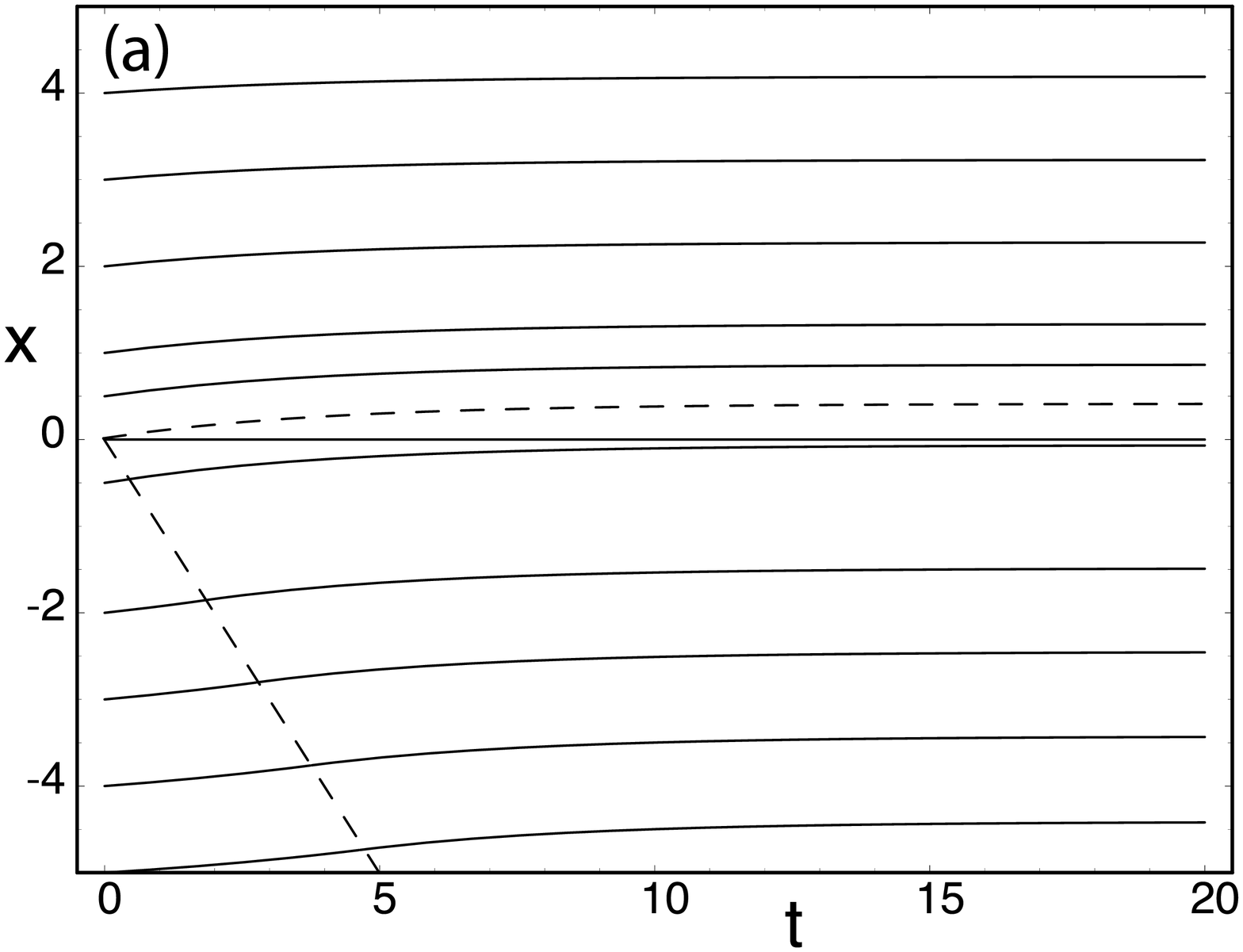} \hfill
  \includegraphics[width=0.48\linewidth]{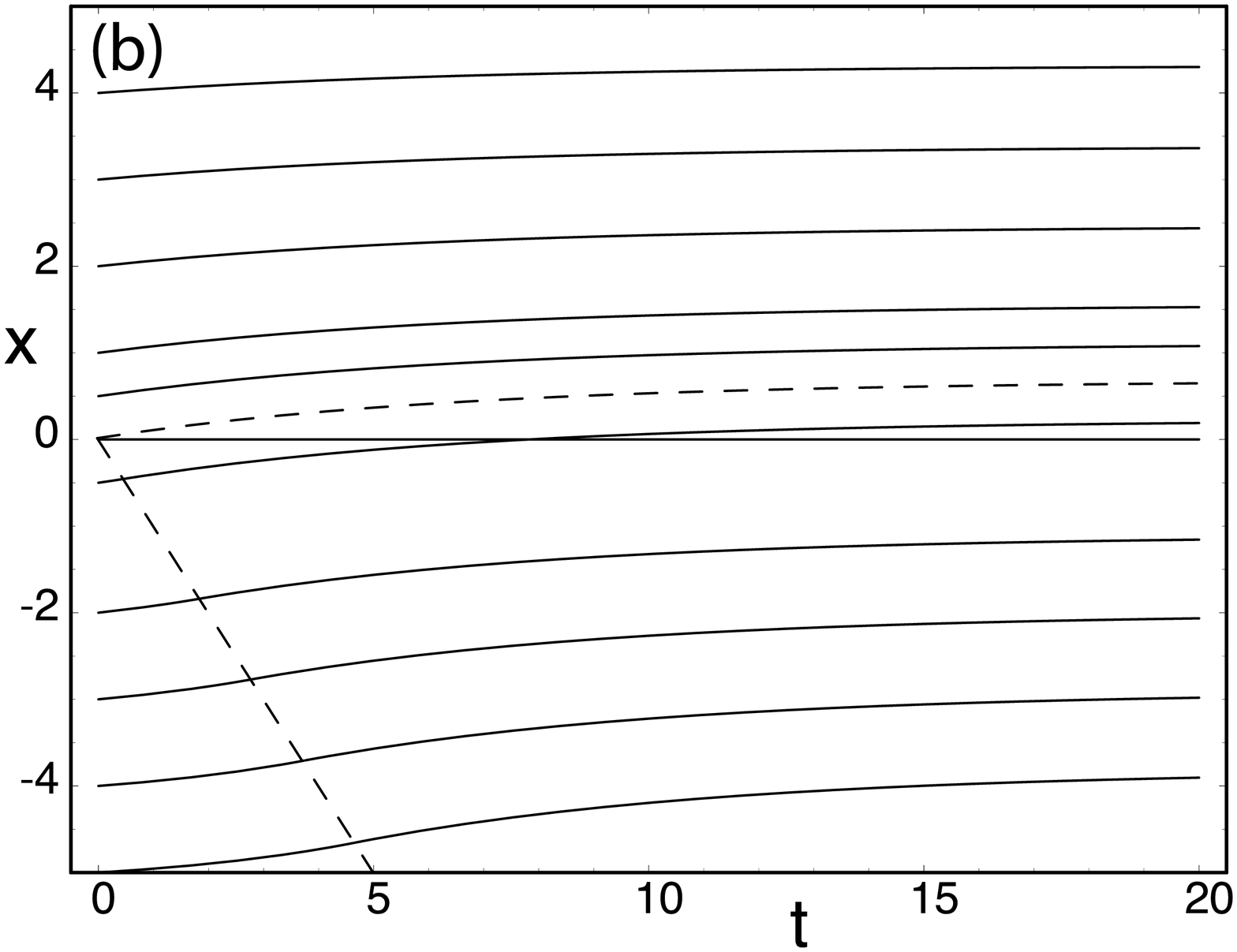} \\[0.5cm]
  \includegraphics[width=0.48\linewidth]{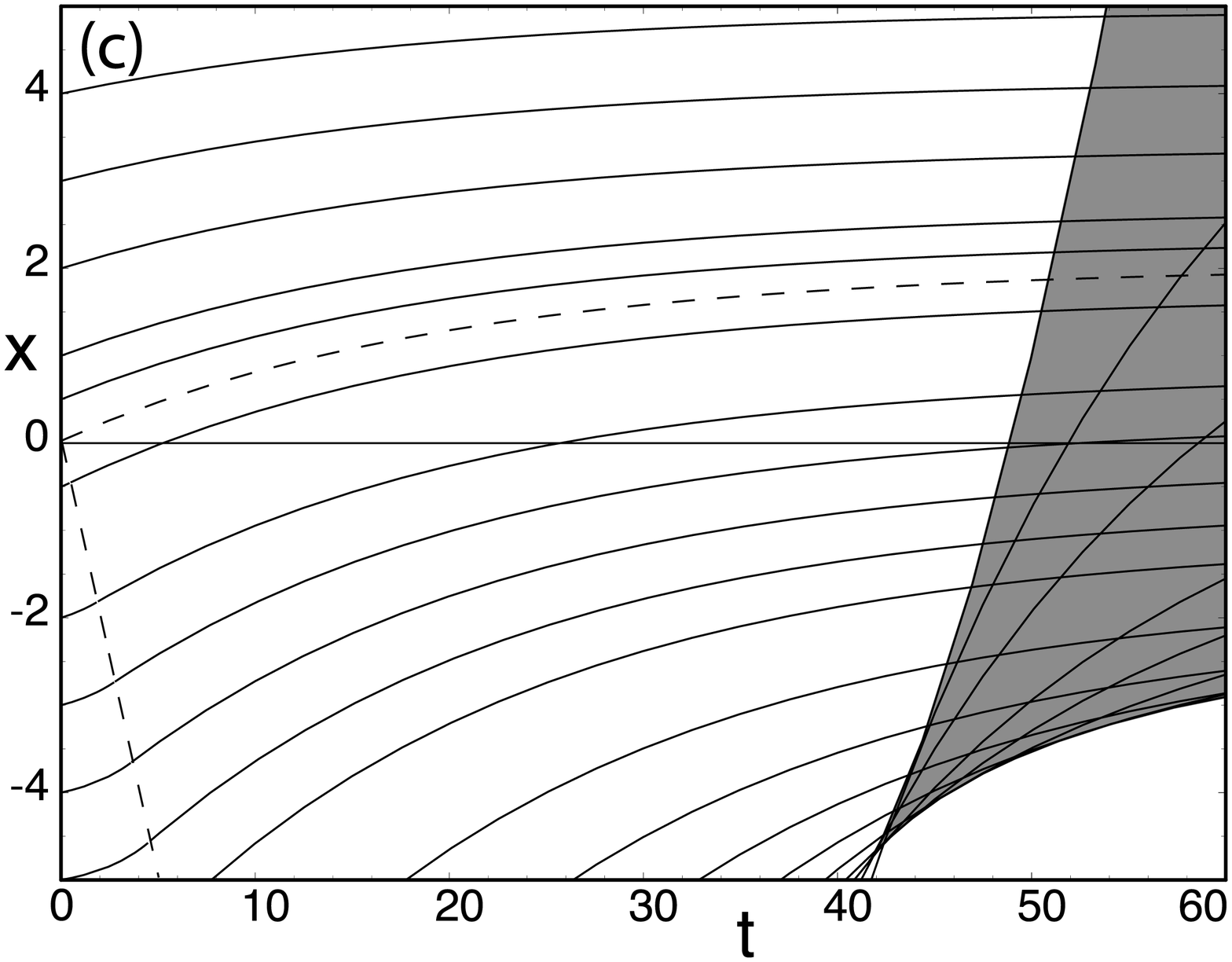} \hfill
  \includegraphics[width=0.48\linewidth]{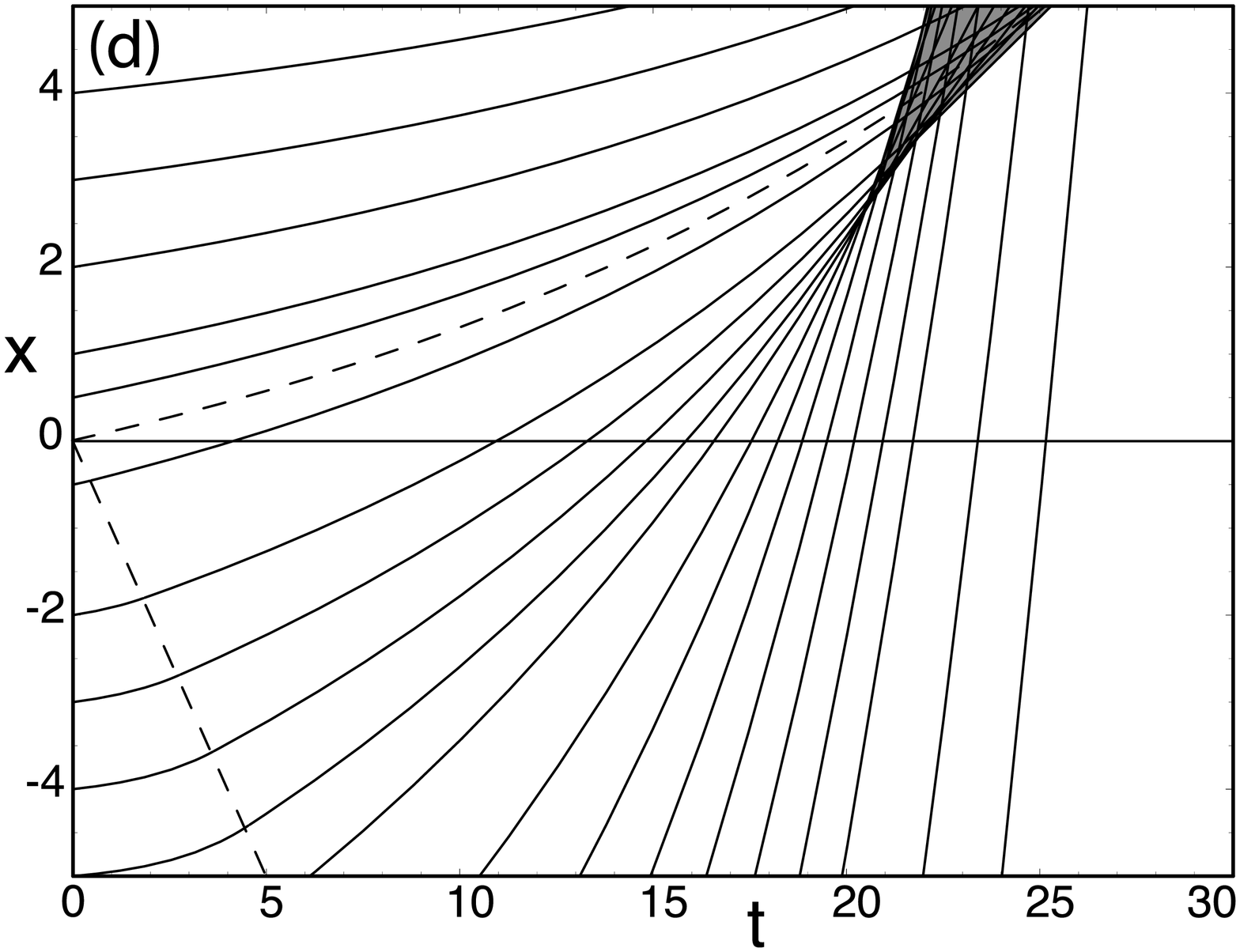}
\caption{Model $\mathcal{P}$: Characteristic curves with constant $\alpha$ 
  for a perturbation $R_0(x)$ of Eq.~(\ref{eq:BCRE-R0}) with
  $r_0=0.1$, $\delta=5.0$ and slopes (a) $m=-0.05$, (b) $m=0.05 <
  1/\delta$, (c) $m=0.15 < 1/\delta$ and (d) $m=0.25 > 1/\delta$.  The
  shaded area indicates the region with shocks. The dashed curves mark
  the boundaries between the different sectors, cf figure
  \ref{fig:sectors}.}
\label{fig:c-curves}
\end{figure}

In sectors (I) and (III) an explicit expression for the profiles can
be obtained. From the integrated version of equation (\ref{eq:BCRE-map:t})
one easily gets the result
\begin{equation}
  \label{eq:BCRE-R-sol-I+III}
  R(t,x)=R_0(\alpha(t,x)) e^{(m\pm 1/\delta)t},
\end{equation}
where $\pm$ refers to sector (I), (III), respectively. Together with
equation (\ref{eq:Z-of-ab}) this result shows that the system has a simple
time evolution along the characteristic curves of constant
$\alpha(t,x)$. Using equations (\ref{eq:Z-of-ab}),
(\ref{eq:BCRE-R-sol-I+III}) and $\beta=-t-x$, we obtain the explicit
expression for the coordinate mapping in sector (I),
\begin{equation}
  \label{eq:BCRE-a(t,x)}
  \frac{1}{m}Z(t,x)=
\alpha(t,x)=x+\frac{1-e^{(m+1/\delta)t}}
{m+(1/\delta) e^{(m+1/\delta)t}} W\left[\frac{r_0}
{m+1/\delta}e^{r_0+x/\delta} \left(m + (1/\delta) 
e^{(m+1/\delta)t}\right)\right].
\end{equation}
The result for sector (III) is obtained from the latter expression by
the replacement $\delta \to -\delta$. equations (\ref{eq:BCRE-R-sol-I+III}),
(\ref{eq:BCRE-a(t,x)}) are our final result for the profiles in
sectors (I) and (III). In sector (II) the characteristic curves
$x_\alpha(t)$ which maps to a constant $\alpha$ can be obtained again
explicitly but we were unable to invert them to obtain $\alpha(t,x)$.
The characteristic curves read
\begin{equation}
  \label{eq:BCRE-cc-II}
  x_\alpha(t)=\frac{1}{m-1/\delta}\left[
h(\alpha)e^{(m-1/\delta)t} + \ln(h(\alpha)/r_0) + m\alpha -r_0
\right]
\end{equation}
with the function
\begin{equation}
  \label{eq:BCRE-h(a)}
  h(\alpha)=W^{\frac{2/\delta}{m+1/\delta}}\left[
 r_0 e^{r_0-m\alpha}\right]
W^{\frac{m-1/\delta}{m+1/\delta}}\left[ 
r_0 e^{r_0+\alpha/\delta}
\right].
\end{equation}
As in sectors (I) and (III) the curves behave exponentially in time
with, however, more complicated amplitudes. The characteristic curves
are shown for $r_0=0.1$, $\delta=5.0$ and different slopes $m$ in
figure \ref{fig:c-curves}.

%
%

With the characteristic curves at hand, the solutions for $R(t,x)$ and
$Z(t,x)$, at a fixed time, are given in parametric form in the $x-R$
or $x-Z$ plane by the curves $[x_\alpha(t), R(\alpha,\beta=-x-t)]$,
$[x_\alpha(t), m\alpha]$, respectively, with $\alpha$ as running
parameter. 

\subsubsection{Appearance of shocks}

Before we discuss the resulting profiles, we will study the
possibility for the occurrence of shocks, i.e., discontinuities of the
profiles which develop at a finite time. Such kind of singularities
are possible for non-linear dynamics since adjacent characteristic
curves can bend differently, leading even for small differences in
curvature at small times to intersecting curves at larger times.
Beyond the time at which the curves cross for the first time, there is
a region where a unique solution no longer exists. Two such situations
are visualized for our model in figures \ref{fig:c-curves}(a) and (c).
If characteristic curves cross each other they form an envelope. The
shock position is then determined by the cusp of the envelope.  While
we leave the precise definition and the calculation of the envelope to
Appendix \ref{sec:app_shock}, we discus here the criterion for the
existence of shocks in our model.  We are interested in not too large
amplitudes $r_0$. For $r_0<\sqrt{2}$, there is a simple condition for
the formation of shocks. Then it can be shown, see Appendix
\ref{sec:app_shock}, that shocks occur only if the slope of the
profile of the resting grains is larger than the critical value
\begin{equation}
\label{eq:BCRE-shock_m}
m_c = \frac{\sqrt{2} - 1}{\delta}.  
\end{equation}
Thus for broad perturbations in the moving phase shocks occur already
for small slopes $m$. The exact time $t_s$ and position of the shock
is in general difficult to calculate. For sufficiently large $m$ ($m >
1/(2r_0\delta)$ if $r_0<1$ and $m>1/(\delta(1+r_0))$ if $r_0>1$) an
exact expression can be obtained and is given in Appendix
\ref{sec:app_shock}. On the other hand, for slopes close the critical
one the shock time diverges as
\begin{equation}
\label{eq:BCRE-shock_sing}
t_s \sim (m-m_c)^{-1}.
\end{equation}
For our above choice of parameters $r_0$ and $\delta$ for
figure \ref{fig:c-curves} the critical slope is $m_c=0.0828$.

\subsubsection{Different avalanche dynamics}

\begin{figure}[t] 
\includegraphics[width=0.48\linewidth]{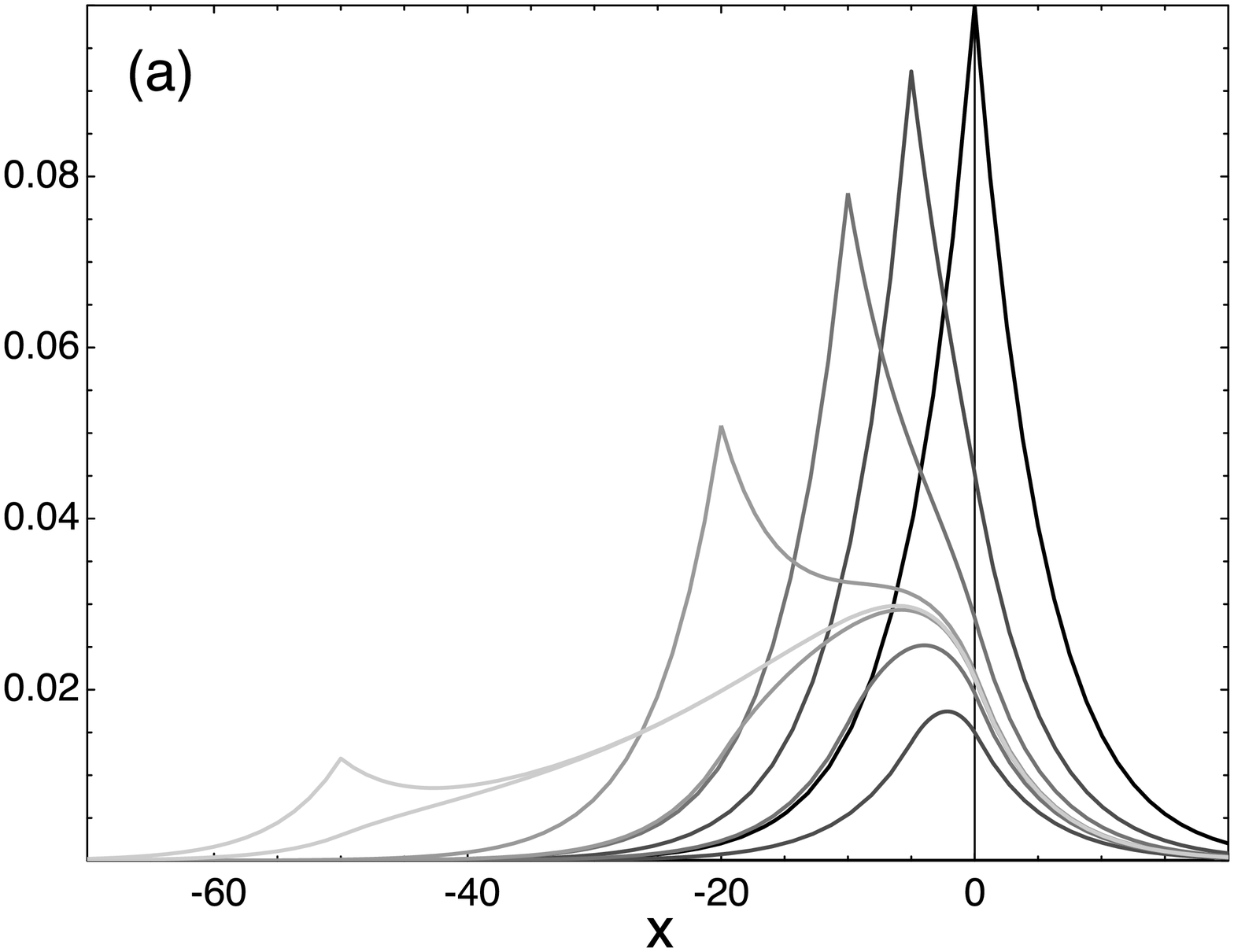} 
\hfill
\includegraphics[width=0.48\linewidth]{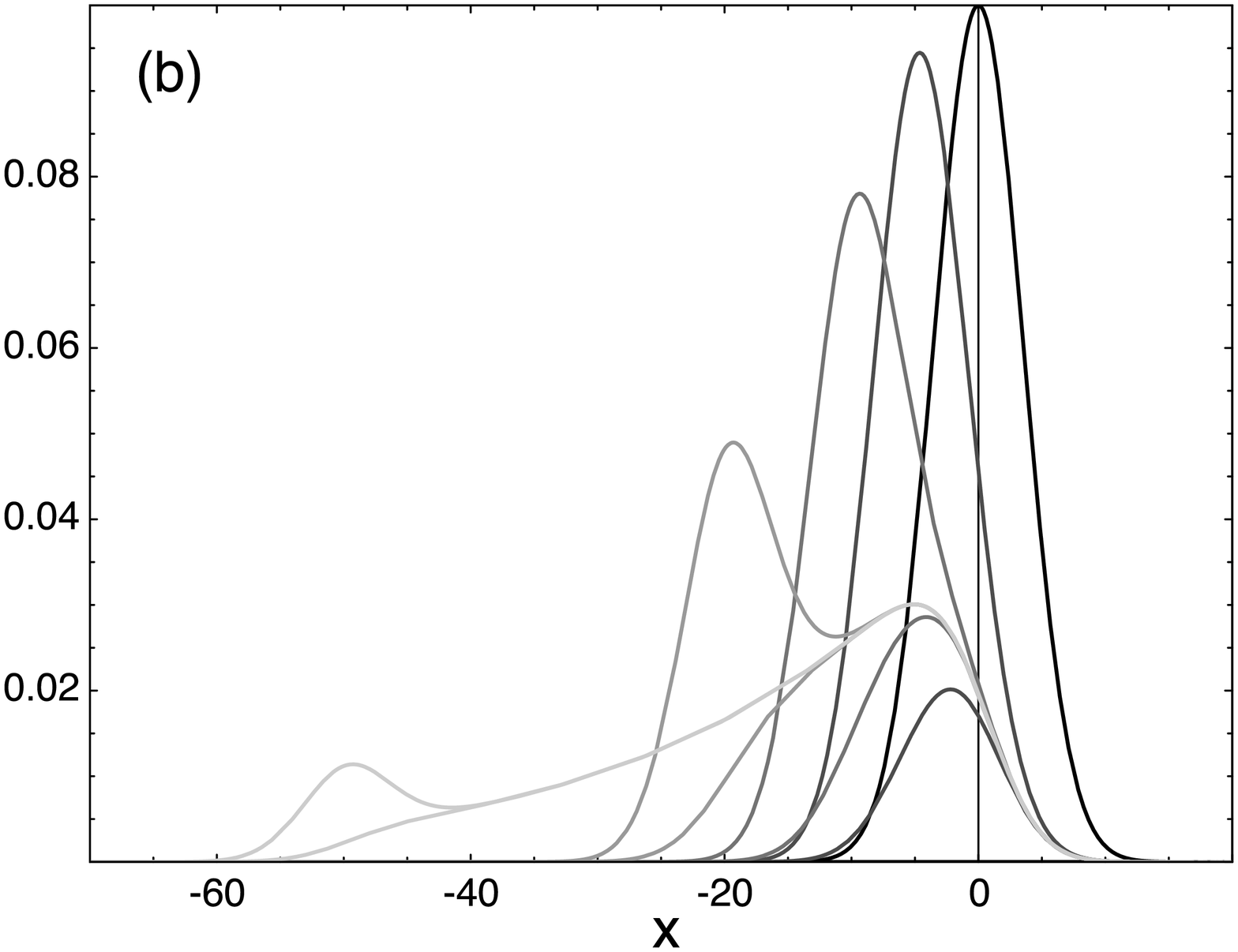} 
\caption{Model $\mathcal{P}$ with $b=0$: Fixed time profiles $Z(t,x)-mx$ and 
  $R(t,x)+Z(t,x)-mx$ so that the gap between two corresponding curves
  represents the thickness of the mobile phase. Plot (a) is for
  $R_0(x)$ of Eq.~(\ref{eq:BCRE-R0}) and (b) for a Gaussian $R_0(x)$,
  both for $m=-0.05$. The profiles are for times $t=0$, $5$, $10$,
  $20$ and $50$.  Here and in the following plots of profiles the
  plotting intensity decreases with increasing time.}
\label{fig:profiles_-0.05}
\end{figure}

\begin{figure}[h!] 
\includegraphics[width=0.47\linewidth]{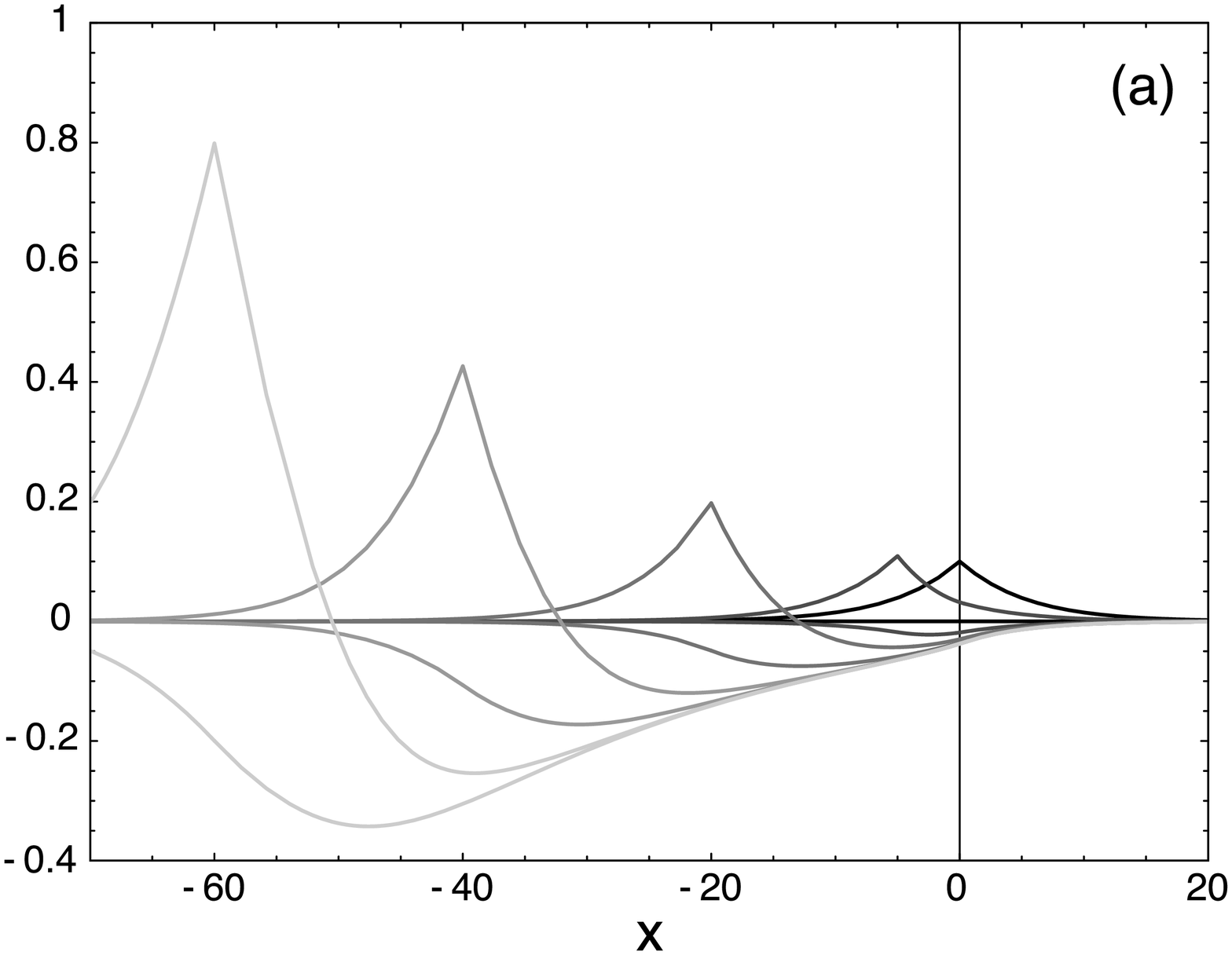} 
\hfill
\includegraphics[width=0.48\linewidth]{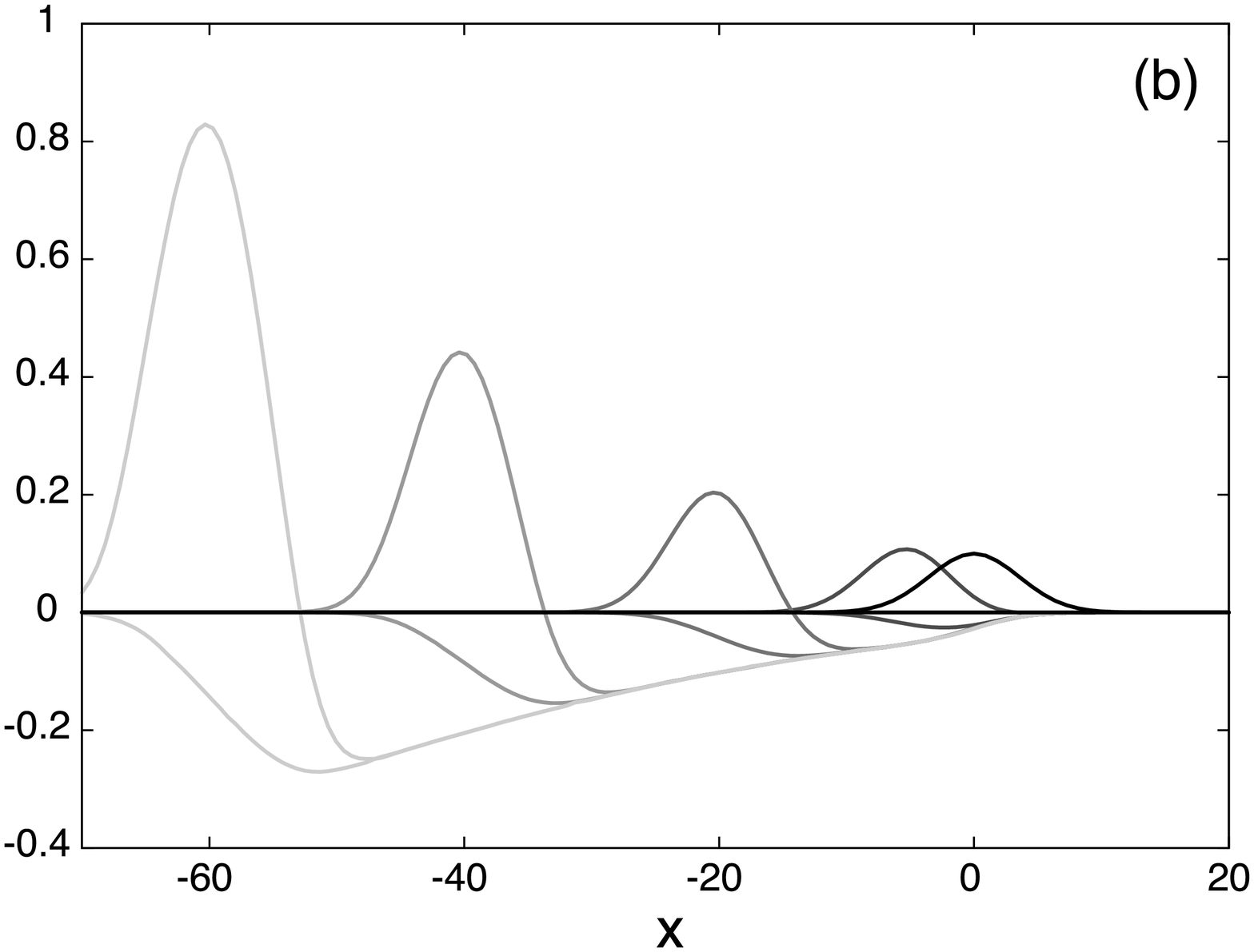} 
\caption{Analog of Fig.~\ref{fig:profiles_-0.05} but for $m=0.05$ and 
times $t=0$, $5$, $20$, $40$ and $60$.}
\label{fig:profiles_0.05}
\end{figure}

\begin{figure}[h!] 
  \includegraphics[width=0.46\linewidth]{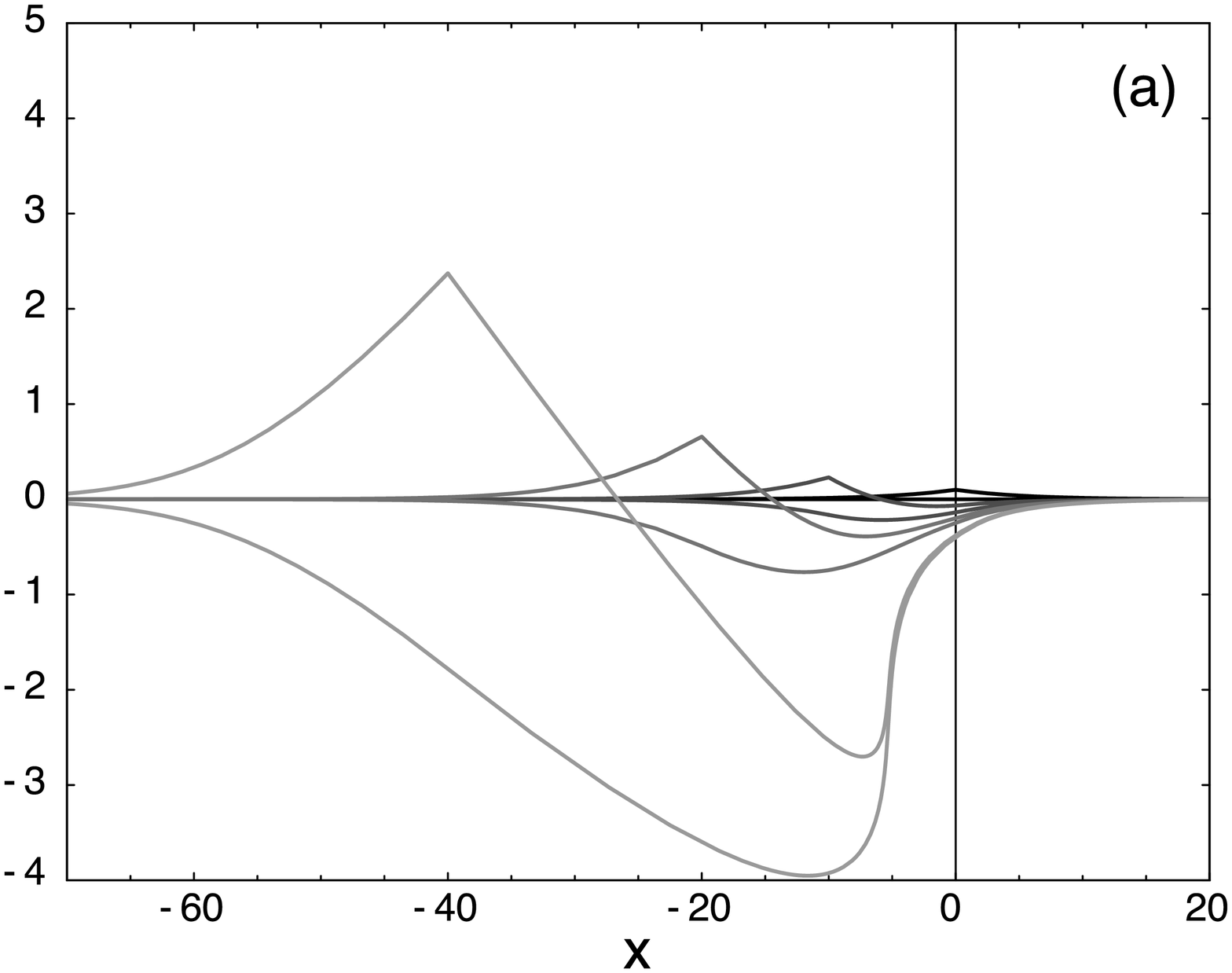} \hfill
  \includegraphics[width=0.48\linewidth]{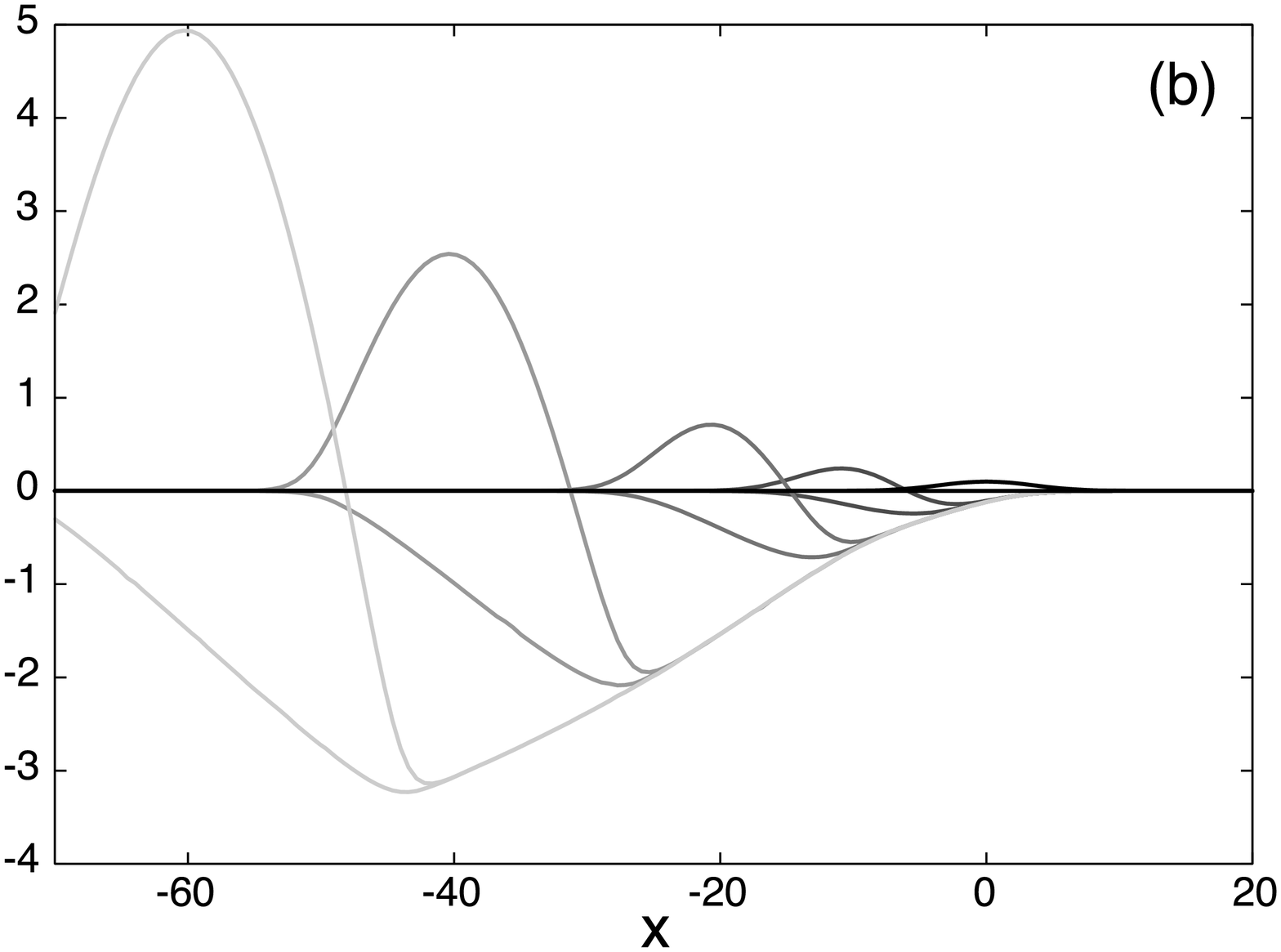}
\caption{Analog of Fig.~\ref{fig:profiles_-0.05} but for $m=0.15$ and 
  times $t=0$, $10$, $20$, $40$ and $60$ for (b) only.  In plot (a) a
  shock occurs at the uphill end of the avalanche (shock time
  $t_s=41.71$). Note that for the Gaussian $R_0(x)$, plot (b), shocks
  are generated only for larger values of $m$, see text.}
\label{fig:profiles_0.15}
\end{figure}

Figure \ref{fig:c-curves} shows four possible situations: (a) A slope
$m$ which is larger than critical $m_c$ but smaller than $1/\delta$ so
that the characteristic curves saturate at large times, (b) a slope
smaller than the critical $m_c$, (c) a slope which is larger than both
the critical $m_c$ and $1/\delta$ so that the curves grow
exponentially in time, and (d) the case of a negative $m$. The
solutions for the corresponding profiles of $R(t,x)$ and $Z(t,x)$ are
shown in parts (a) of figures
\ref{fig:profiles_-0.05}~-~\ref{fig:profiles_0.15}.  Plotted are the
two curves $Z(t,x)-mx$ (always corresponding to the lower curve) and
$Z(t,x)-mx+R(t,x)$ so that the gap between the two curves represents
the layer of moving grains. The maximum of the perturbation propagates
downhill with a constant velocity which is $1$ in our rescaled units.
For a negative slope $m$ (corresponding to an actual slope
smaller than the angle of repose) all grains of the perturbation come
finally to rest, generating a bump on the initial static profile which
corresponds to the baseline in the plot, see figure
\ref{fig:profiles_-0.05}(a). The amplitude of the perturbation decays
exponentially at large times, $R(t,x=-t)\simeq r_0 \exp(r_0+mt)$. If
we define the downhill end of the bump by the condition that the
maximum of the perturbation in $R(t,x)$ has decayed to some fraction
$\epsilon \ll 1$ of the initial amplitude, then the width of the bump
scales like $\ln(\epsilon)/m$ for small $r_0$ since the peak in
$R(t,x)$ moves with a velocity of one. Thus the final width of the
deposited amount of grains is independent of the amplitude and width
of the perturbation but only determined by the initial slope of the
static phase. For positive slopes $m$ the amplitude $R_\text{max}$ of
the perturbation grows at large times linear,
\begin{equation}
\label{eq:bcre_R_max}
R_\text{max} \simeq m(1+m \delta)\, t, 
\end{equation}
with a growth rate which is independent of the initial amplitude
$r_0$.  However, there is a broad transient behavior with logarithmic
corrections at intermediate time scales. 
The exponential time behavior found at the beginning of this section
for an initially homogeneous mobile phase is recovered for $m<0$ but
is in contrast to the linear growth for positive $m$.

For both positive and negative initial slopes, the profile $Z(t,x)$ of
the static phase does no longer evolve after the avalanche has passed.
The resulting profile $Z_\infty(x)$ becomes thus time independent for
times larger than a (position dependent) time scale.  This asymptotic
profile is only well defined if no shock occurs. In the latter case it
is implicitly given by
\begin{equation}
\label{eq:BCRE_Z_asympt}
\frac{1}{m-1/\delta}\left[ \ln \left( \frac{h(Z_\infty(x)/m)}{r_0} \right)
+Z_\infty(x)-r_0 \right] = x,
\end{equation}
where the function $h$ is given by equation (\ref{eq:BCRE-h(a)}). The latter
expression is valid in sector (II) which is the relevant region for
large times, see figure \ref{fig:sectors}. From this result one can
obtain the slope $\partial_x Z_\infty$ at a given value of $Z$,
\begin{equation}
\label{eq:BCRE_Z_asympt_slope}
\partial_x Z_\infty (x) = \left( m - 1/\delta \right) 
\left[ \frac{h'(Z_\infty(x)/m)}{m h(Z_\infty(x)/m)} + 1 \right]^{-1}.
\end{equation}
The behavior of $Z_\infty$ depends on the sign of $m$. For negative
$m$ the deposited grains form a bump that was described above and
whose exact shape is given by equation (\ref{eq:BCRE_Z_asympt}). Across this
bump, we find that the change of the slope as compared to the initial
slope $m$ is always rather small. Far away from the bump (at large
positive and negative $x$) one has $h'/h \simeq - 1/\delta$, and thus
$\partial_x Z_\infty =m$ remains, of course, unchanged from the
initial profile.  For positive $m$ the profile $Z(t,x)$ shows again a
constant slope after the avalanche has passed, i.e., $Z_\infty =
m_\infty x$, cf figure \ref{fig:profiles_0.05}. The asymptotic slope can
be obtained from the behavior of the function $h$ at large negative
arguments. We find
\begin{equation}
\label{eq:bcre_asymp_slope}
m_\infty = m \left[ 1 + \left(\frac{2m}{m\delta+1+\sqrt{2}}\right) 
\left( \frac{1}{m_c-m} \right) \right].
\end{equation}
Interestingly, the relative change of the slope is independent of the
amplitude $r_0$ of the perturbation and depends only on the product of
the initial slope $m$ and the width $\delta$ of the perturbation.
Surprisingly, the expression in the square brackets in equation
(\ref{eq:BCRE_Z_asympt_slope}) is larger than one for positive $m$,
and even diverges as the slope $m$ approaches the critical value $m_c$
beyond which shocks occur, $m_\infty \sim (m_c-m)^{-1}$. It
is important to note that the layer of moving grains decays to zero at large 
times at the uphill end of the avalanche, although the slope of the static layer
is steeper than before the avalanche started.  This behavior can be
easily understood from the property of equation (\ref{eq:SVUconstant}) in
which the exchange between the static and rolling phase is
proportional to $R$. Of course, physically, there will be a maximum angle for 
the stability (even for $R=0$) beyond which the above predictions are irrelevant, and
an extended model has to be considered. 

An important quantity is the total size of the avalanche. Within our
model, we define the size $I(t)$ of an avalanche as the spatially
integrated amount of mobile grains, i.e., 
\begin{equation}
\label{eq:def_size}
I(t)=\int_{-\infty}^\infty R(t,x) \, dx \, .
\end{equation}
The integration can be performed for each sector separately by a
change of variables to the characteristic coordinate $\alpha$. The
contribution from sector (III) can be neglected at larger times since
the avalanche starts at $x=0$ to propagate to the left (downhill), cf
figure \ref{fig:sectors}.  For negative $m$, the perturbation decays
exponentially, and thus we obtain for the size  
\begin{equation}
  \label{eq:bcre_m_neg_size}
  I(t)=2 r_0 \delta \, e^{r_0+mt}.
\end{equation}
For positive $m$ we observe that the size of the avalanche shows a
quadratic increase in time,
\begin{equation}
  \label{eq:bcre_m_pos_size}
  I(t)= m^2 \delta \, \frac{1+m\delta}{1-m\delta} \, t^2,
\end{equation}
at asymptotically large times. Interestingly, the growth of the
avalanche depends only on its initial width $\delta$ but not on the
amplitude $r_0$. By comparison with the scaling of the amplitude of
the avalanche, cf equation (\ref{eq:bcre_R_max}), we observe that the
width of the avalanche must grow also linear in time $\sim
m\delta/(1-m\delta) t$.

So far we have studied mainly an initial perturbation which is
particularly suited for obtaining analytical results. In order to
check the robustness of our results with respect to the precise form
of $R_0(x)$ we have chosen also a Gaussian $R_0(x)=r_0
\exp(-x^2/\delta^2)$ together with the same $Z_0(x)=mx$ as before.
Contrarily to the previous case, the initial perturbation has no cusp
at $x=0$. By a numerical computation of the integral of equation
(\ref{eq:BCRE-map:t}) we obtained the profiles shown in parts (b) of
figures \ref{fig:profiles_-0.05}~-~\ref{fig:profiles_0.15}, using
$r_0=0.1$, $\delta=5$ as before. As can be observed from the plots the
characteristic features can be regarded as robust.  However, the
moving layer, i.e., the gap between the upper and lower graph decays
more rapidly due to the faster decay of the Gaussian profile.  Of
course, the critical slope $m_c$ for shocks is no longer given by
equation (\ref{eq:BCRE-shock_m}). We observe that for the Gaussian
$R_0(x)$ shocks occur only beyond a $m_c$ which is {\em increased}
compared to the exponentially decaying profile of equation
(\ref{eq:BCRE-R0}) with the same width at half height, 
cf figure \ref{fig:profiles_0.15}. 

%


\subsection{Strong stress anisotropy: small $b$}
\label{sec:constant_b_finite}

\subsubsection{Analytical result for general slopes $m$}

\begin{figure}[t] 
\includegraphics[width=0.5\linewidth]{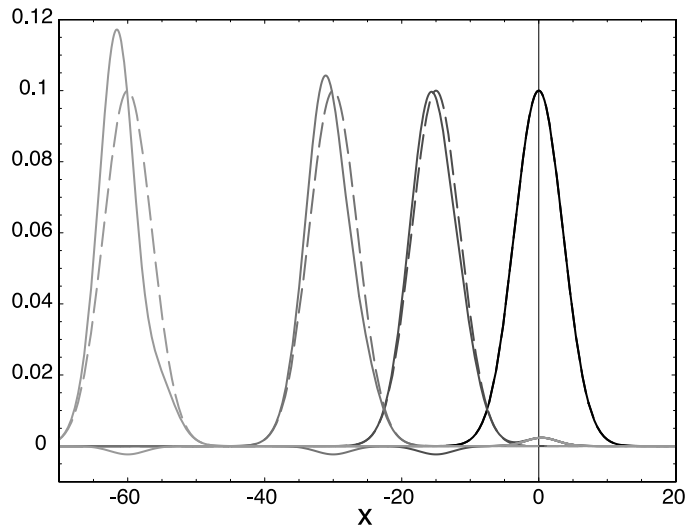} 
\caption{Model $\mathcal{P}$ at the angle of repose ($m=0$) for
  finite horizontal stress, $b=0.5$: Profiles for a Gaussian $R_0(x)$
  at times $t=0$, $15$, $30$ and $60$. The dashed curves correspond to
  the absence of horizontal stress, $b=0$, where the perturbation
  $R_0$ propagates without changing its shape.}
\label{fig:profiles_Gauss_m0_b05}
\end{figure}

So far we have assumed such a strong stress anisotropy that there is
no horizontal stress $\sigma_{xx}$, i.e., $b=0$ in the model of
equation (\ref{eq:SVUconstant}). In this section we will study the
influence of a small $\sigma_{xx}$ on the avalanche dynamics we found
in the previous section. Although steady state simulations suggest a
value of $b$ close to one \cite{SEGHLP01,dC04}, it is interesting to
study the regime of small $b$ in order to compare to the BCRE model.
The method of characteristics can be applied of course to arbitrary
values of $b$, yielding a system of equations which we could not solve
explicitly so that we had to resort to a perturbative treatment.
Thus, we consider the terms proportional to $b$ in equation
(\ref{eq:SVUconstant}) as small perturbations of the $b=0$ solution.
This can be done by the following ansatz
\begin{subequations}
  \label{eq:bcre_ansatz_finite_b}
\begin{eqnarray}
  Z &=& Z_1 + b Z_2 \\
  R &=& R_1 + b R_2,
\end{eqnarray}
\end{subequations}
where $Z_1$ and $R_1$ denote the solution for $b=0$ of the previous
section. Although one expects realistic values of $b$ of order unity,
the perturbative calculation should allow for a qualitative assessment
of the effect of a finite horizontal stress.  The dynamics of the
corrections are then described by the following {\it linear} coupled
equations
\begin{subequations}
  \label{eq:bcre_2nd_order_systen}
\begin{eqnarray}
  \label{eq:bcre_2nd_order_systen_Z}
\dr_t Z_2 &=& - R_1 \left( \dr_x Z_2 + \dr_x R_1 \right) -R_2 \dr_x Z_1 \\
  \label{eq:bcre_2nd_order_systen_R}
\dr_t R_2 &=& \dr_x R_2 + R_2 \dr_x Z_1 + R_1 
\left( \dr_x Z_2 + \dr_x R_1 \right)\, .
\end{eqnarray}
\end{subequations}
%
\begin{figure}[t] 
\includegraphics[width=0.48\linewidth]{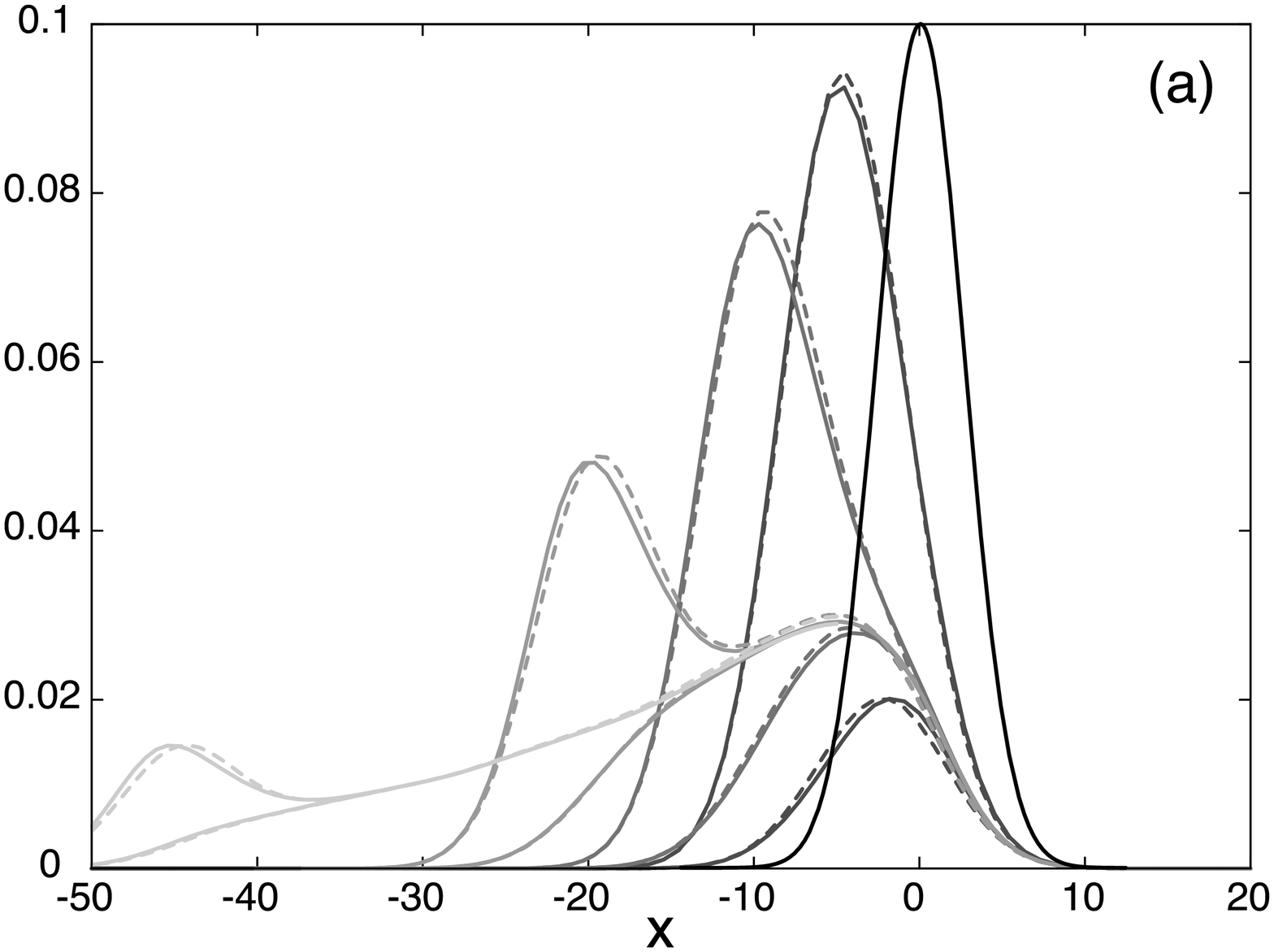}
\hfill
\includegraphics[width=0.48\linewidth]{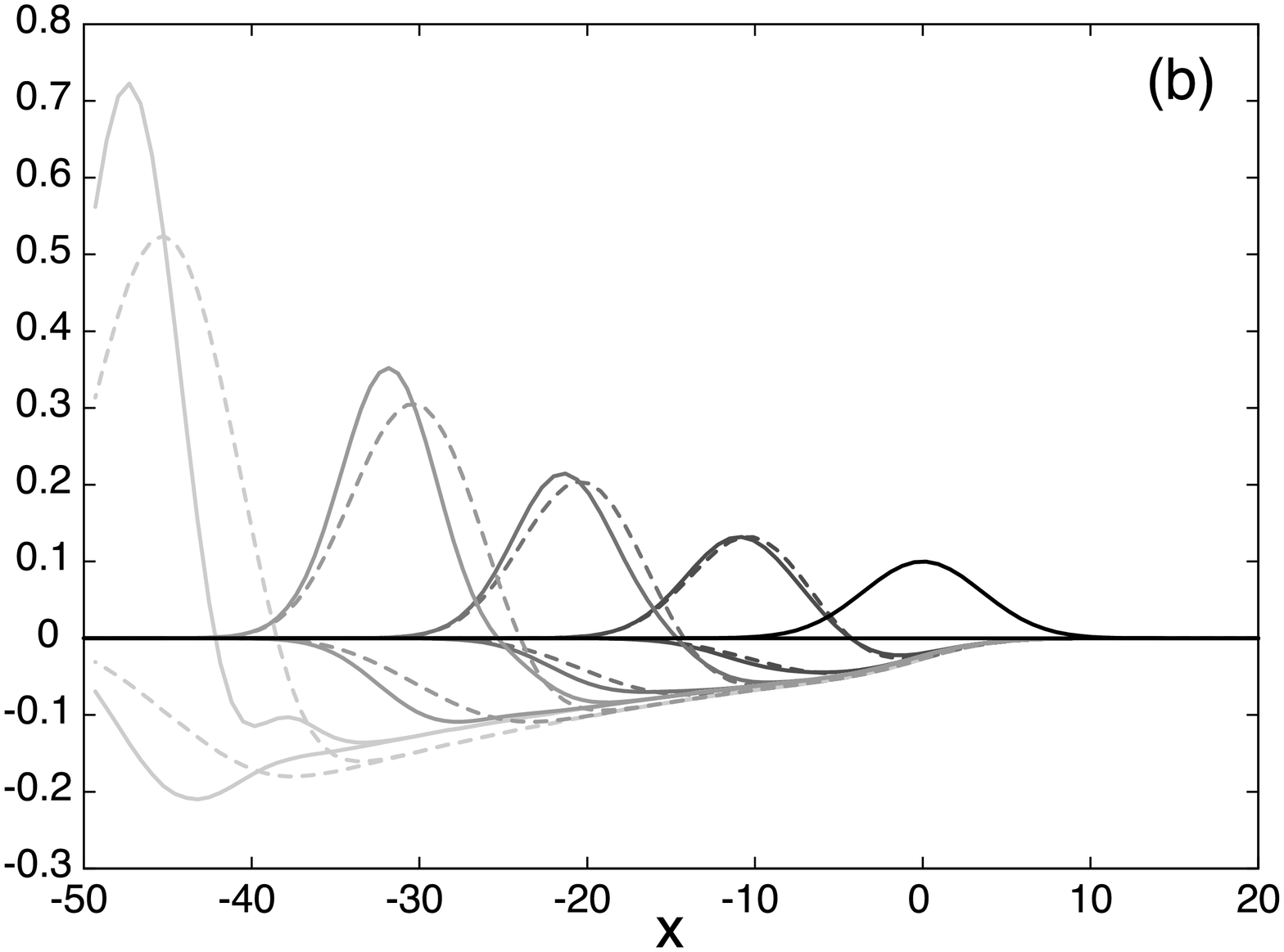}
\caption{Model $\mathcal{P}$ with finite $b=0.5$: Profiles for a Gaussian 
  $R_0(x)$ and (a) $m=-0.05$ (for $t=0$, $5$, $10$, $20$ and $45$),
  (b) $m=0.05$ (for $t=0$, $10$, $20$, $30$ and $45$). The dashed
  curves represent the corresponding profiles for $b=0$ of section
  \ref{sec:constant_b_zero} and are shown for comparison.}
\label{fig:profiles_finte-b_-0.05_0.05}
\end{figure}
%
Since we consider corrections of linear order in $b$, the terms
containing derivatives of the profiles $R_2$ and $Z_2$ have exactly
the same form as those in equation (\ref{eq:SVUconstant}) with $b=0$. Thus,
the characteristic directions $\zeta_+= -1$, $\zeta_-=R_1$ and
characteristic curves remain unchanged for small $b$.  In terms of the
characteristic coordinates $\alpha$ and $\beta$, the corrections obey
the equations
\begin{subequations}
\label{eq:bcre_fibite_b_charac_eqs}
\begin{eqnarray}
  \label{eq:bcre_fibite_b_charac_eqs_a}
\dr_\alpha Z_2 + \frac{1+R_1}{R_1} \dr_\alpha R_2 -
\left(\frac{R_2}{R_1} \dr_x Z_1 + \dr_x R_1 \right) \dr_\alpha t & = & 0\\
  \label{eq:bcre_fibite_b_charac_eqs_b}
\dr_\beta Z_2 +\left( R_2 \dr_x Z_1 + R_1 \dr_x R_1 \right) \dr_\beta t 
& = & 0.  
\end{eqnarray}
\end{subequations} 
Here all functions have to be considered as depending on $\alpha$ and
$\beta$, in particular $t(\alpha,\beta)$ is given by
equation (\ref{eq:BCRE-map:t}).  In order to express the derivatives with
respect to $x$ as functions of $\alpha$, $\beta$, we use $\dr_x =
\dr_x \alpha \, \dr_\alpha + \dr_x \beta \, \dr_\beta$ together the
relations
\begin{equation}
  \label{eq:bcre_map_relations}
  \dr_x \beta = -1, \quad 
\dr_x \alpha = \frac{1}{1+R_1} \frac{1}{\dr_\alpha x}
\end{equation}
with $x(\alpha,\beta)$ given by equation (\ref{eq:BCRE-map:x}). Using these
relations and the solutions for $b=0$ of
equation (\ref{eq:BCRE-b=0-solution}) with the initial condition
$Z_0(\alpha)=m\alpha$, the equation (\ref{eq:bcre_fibite_b_charac_eqs}) can
be rewritten after integration over $\alpha$ as
\begin{subequations}
\label{eq:bcre_finite_b_final}
\begin{eqnarray}
\label{eq:bcre_finite_b_final_I}
Z_2+\frac{1+R_1}{R_1} \, R_2 + \ln \left( \frac{1+R_1}{1+R_0(-\beta)} \right)
-\int_{-\beta}^\alpha d \alpha' \dr_\beta R_1(\alpha',\beta) \dr_{\alpha'}
x(\alpha',\beta) & = & 0 \\
\label{eq:bcre_finite_b_final_II}
\dr_\beta Z_2 + \frac{R_1}{1+R_1} \dr_\beta R_1 + \frac{m}{(1+R_1)^2 
\dr_\alpha x} \left( \frac{R_1^2}{1+R_1} - R_2 \right) & = & 0,
\end{eqnarray}
\end{subequations}
where the functions depend on $\alpha$ and $\beta$ unless arguments
are written explicitly. The explicit expression for $\dr_\alpha x$
can be obtained from the solution for $b=0$, leading to
\begin{equation}
  \label{eq:bcre_finite_b_dr_a_x}
  \dr_\alpha x = \frac{1}{1+R_0(\alpha)} - \int_{-\alpha}^\beta d\beta'
\frac{\dr_\alpha R_1(\alpha,\beta')}{[1+R_1(\alpha,\beta')]^2}.
\end{equation}
The function $R_2$ can be eliminated from equation
(\ref{eq:bcre_finite_b_final_II}) by using
equation (\ref{eq:bcre_finite_b_final_I}), and the resulting linear ordinary
differential equation for $Z_2$ can be integrated easily. The result
is
\begin{equation}
  \label{eq:bcre_result_Z_2}
  Z_2(\alpha,\beta) = 
\exp\left[ \int_{-\alpha}^\beta d\beta' g_1(\alpha,\beta') \right]
\int_{-\alpha}^\beta d\beta' g_2(\alpha,\beta') \exp\left[
-\int_{-\alpha}^{\beta'} d\beta'' g_1(\alpha,\beta'') \right],
\end{equation}
with the functions
\begin{subequations}
\label{eq:bcre_aux-fcts}
\begin{eqnarray}
\label{eq:bcre_aux-fcts-g}
g_1(\alpha,\beta) & = & -\frac{m R_1}{(1+R_1)^3 \dr_\alpha x} \\ 
\label{eq:bcre_aux-fcts-h}
g_2(\alpha,\beta) & = & - \frac{R_1}{1+R_1} \dr_\beta R_1 -
\frac{mR_1}{(1+R_1)^3 \dr_\alpha x} \left\{
R_1 + \ln \left( \frac{1+R_1}{1+R_0(-\beta)} \right)
- \int_{-\beta}^\alpha d\alpha' \dr_\beta R_1(\alpha',\beta) 
\dr_{\alpha'} x(\alpha',\beta) 
\right\}.
\end{eqnarray}
\end{subequations}
This is our final result for $Z_2$, the profile $R_2$ can be computed
now from equation (\ref{eq:bcre_finite_b_final_I}). Using the explicit
result for $R_1$ of equation (\ref{eq:BCRE-b=0-solution-R}) the
multiple integrals can be performed easily numerically. The resulting
profiles are shown in figures \ref{fig:profiles_finte-b_-0.05_0.05}
for the parameters $r_0$, $\delta$ and $m$ of section
\ref{sec:constant_b_zero} with $b=0.5$. For comparison the solution
for $b=0$ are also shown as dashed curves. As can be seen from figure
\ref{fig:profiles_finte-b_-0.05_0.05}(a), for negative $m$ the
avalanche and the static profile are not much effected by the presence
of horizontal stress $\sim b$. In contrast, for positive slopes $m$
there is more visible effect of the horizontal stress.  This effect
grows with increasing time scale, cf figure
\ref{fig:profiles_finte-b_-0.05_0.05}(b). As one could naively expect,
horizontal stress has the tendency to shift the peak in dynamics phase
downhill (to the left). This shift will be analyzed quantitatively
below for the case $m=0$ and by comparing the numerical results of
figures
\ref{fig:profiles_finte-b_-0.05_0.05}-~\ref{fig:profiles_Gauss_m0_b05}
the shift appears to be independent of $m$. For the static profile
$Z(t,x)$ the horizontal stress leaves the final slope after the
avalanche almost unchanged but it produces a steeper slope in $Z(t,x)$
where the moving layer $R(t,x)$ has maximal thickness. There is also a
tendency for the static profile to form a local dell at the peak of
the avalanche, see figure \ref{fig:profiles_finte-b_-0.05_0.05}(b).
This observation becomes especially pronounced for $m=0$, see figure
\ref{fig:profiles_Gauss_m0_b05}.

%
%

\subsubsection{Explicit results at the angle of repose ($m=0$)}

It is obvious form the structure of equations
(\ref{eq:bcre_result_Z_2})-(\ref{eq:bcre_aux-fcts-h}) that major
simplifications occur if the initial static profile is exactly at the
angle of repose, i.e., $m=0$. Thus, we can study easily the effect of
a horizontal stress (finite $b$) in this situation. Let us first
summarize the results in the absence of horizontal stress ($b=0$) when
$m=0$.  The static layer stays for all times at the angle of repose,
$Z_1(t,x)=0$, and the initial perturbation $R_0(x)$ in the moving
layer simply propagates downhill, $R_1(t,x)=R_0(t+x)$. Thus the amount
of grains in both the static and the moving layer are conserved. The
situation is different for finite $b$. The layers of static and
dynamic grains, respectively, have coupled dynamics with the
corrections from finite $b$ given by
\begin{subequations}
\label{eq:bcre_b-finite_m-zero-profiles}
\begin{eqnarray}
\label{eq:bcre_b-finite_m-zero-profiles-Z}
Z_2(\alpha,\beta) & = & R_0(\alpha) - R_0(-\beta) + \ln
\left( \frac{1+R_0(-\beta)}{1+R_0(\alpha)} \right) \\
\label{eq:bcre_b-finite_m-zero-profiles-R}
R_2(\alpha,\beta) & = & \frac{R_0(-\beta)}{1+R_0(-\beta)}
\left\{ R'_0(-\beta) t(\alpha,\beta) - Z_2(\alpha,\beta) 
\right\} \, ,
\end{eqnarray}
\end{subequations}
where the mapping between $t$, $x$ and $\alpha$, $\beta$ has for
$m=0$ the simple form
\begin{subequations}
\label{eq:bcre_b-finite_m-zero-map}
\begin{eqnarray}
\label{eq:bcre_b-finite_m-zero-map-t}
t(\alpha,\beta) & = & - \int_{-\alpha}^\beta 
\frac{d\beta'}{1+R_0(-\beta')} \\
\label{eq:bcre_b-finite_m-zero-map-x}
x(\alpha,\beta) & = & -t(\alpha,\beta) - \beta \, .
\end{eqnarray}
\end{subequations}
These equations provide a closed parametric form of the dynamics for a
general initial perturbation $R_0(x)$. The resulting time evolution of
a Gaussian perturbation $R_0(x)=r_0 \exp(-x^2/\delta^2)$ is shown in
figure \ref{fig:profiles_Gauss_m0_b05}. Compared to the absence of
horizontal stress, $b=0$, there are a number of interesting novel
features. The avalanche amplitude increases and the maximum is shifted
downhill. The layer of static grains displays a bump at the initial
position of the perturbation and a dell which propagates downhill
close to the maximum of the avalanche peak. At large times,
equation (\ref{eq:bcre_b-finite_m-zero-map-t}) leads to $\alpha \simeq x$
which together with $\beta=-x-t$ and
equation (\ref{eq:bcre_b-finite_m-zero-profiles}) yields an explicit
expression for the profiles. For a small avalanche amplitude $r_0$,
the position of the peak follows $x=-t-\delta/2$ for a Gaussian
$R_0(x)$. The maximum of $R(t,x)$ grows linearly with time, $R_\text{max}
\sim b (r_0^2 / \delta ) t$, while in the absence of horizontal stress
($b=0$) it remains constant. Notice that this linear growth was
observed for $b=0$ only at angles larger than the repose angle
($m>0$), cf equation (\ref{eq:bcre_R_max}). The form of the static profile
$Z(t,x)$ can be directly obtained from
equation (\ref{eq:bcre_b-finite_m-zero-profiles-Z}). There are two identical
contributions which are shifted relative to each other by $t$. The first 
contribution is approximately given by $b[ R_0(x) - \ln(1+R_0(x))]$.  This
term represents the bump at the start position of the avalanche. The second 
contribution has $x$ replaced by $x+t$, corresponding to the
dell traveling with the avalanche downhill,
cf figure \ref{fig:profiles_Gauss_m0_b05}. Thus, both structures in the
static layer are determined by the initial profile of the perturbation
in the moving layer.

\section{Linear velocity profile (Model $\mathcal{L}$)}
\label{sec:linear_profile}

It has been argued that a constant velocity profile, as assumed in the
previous Section, is only applicable to thin surface flow
\cite{ARdG99}. For a thicker layer of rolling grains, the velocity
should depend on the amount of mobile grains. Experiments and
simulations for steady deep systems suggest a linear profile for the
average horizontal velocity $u(x,y)=\gamma y$ of the flow.  With this
profile we have $U=\frac{1}{2}\gamma R$, $W=\frac{1}{3}\gamma^2 R^2 =
\frac{4}{3} U^2$ and the conservation conditions of equations
(\ref{conservmatter}) and (\ref{conservmomentum1}) yield the model
\begin{subequations}
\label{SVUlinearfinal}
\begin{eqnarray}
\label{SVUlinearfinalZ}
\dr_t Z & = & -\dr_x Z - b \, \dr_x R,\\
\label{SVUlinearfinalR}
\dr_t R & = & R \, \dr_x R + \dr_x Z + b \, \dr_x R.  
\end{eqnarray} 
\end{subequations}
Here all lengths are divided by $g/\gamma^2$, and time is divided by
$\gamma$. Again, $Z$ is replaced by $Z + \mu x$.  The model contains
after this rescaling only one free parameter, $b$.  It is rather
important to note that the latter model is valid only as long as $R$
remains positive since we obtained it from equation
(\ref{conservmomentum1}) after division by $R$.  Thus the actual
solution of equation (\ref{SVUlinearfinal}) is given by the maximum of
$R=0$ and the formal solution for $R$ of equations
(\ref{SVUlinearfinalZ}) and (\ref{SVUlinearfinalR}). In this Section
we will study the consequences of a $R$ dependent linear velocity
profile both in the absence ($b=0$) and presence ($b \neq 0$) of
horizontal stress.  We note that for an initially uniform amount of
rolling grains $R_0(x)=\varrho$ and a static sand bed with constant
slope $Z_0(x)=mx$ the solution to Eqs.~(\ref{SVUlinearfinal}) is
rather simple. As opposed to the exponential growth for model ${\cal
  P}$, the thickness of the mobile layer increases here only linear in
time, $R(t,x)=\varrho + mt$ and $Z(t,x)= m(x-t)$ decreases
accordingly. As for model ${\cal P}$ this solution is independent of
$b$.

\subsection{Infinite stress anisotropy ($b=0$)}
\label{sec:linear_b_zero}

\subsubsection{Analytical solution}

In the limit of $b=0$ the equations (\ref{SVUlinearfinalZ}) and
(\ref{SVUlinearfinalR}) are decoupled. Such set of equations has been
studied by de Gennes et al.  to describe a thick flow of granular
matter in a bounded geometry \cite{ARdG99}. Here we consider this
simple model in an {\it unrestricted} geometry but we allow for general
initial profiles $R_0(x)$ and $Z_0(x)$. Following again the approach
outlined in Appendix \ref{sec:app_cc}, we obtain the characteristic
equations
\begin{subequations}
\label{eq:lin_b_0_ceqs}
\begin{eqnarray}
\label{eq:lin_b_0_characcoord+}
\dr_\al x - \zeta_+ \dr_\al t & = & 0,\\
\label{eq:lin_b_0_characcoord-}
\dr_\bt x - \zeta_- \dr_\bt t & = & 0,\\
\label{eq:lin_b_0_characSV+}
- \dr_\al Z + (\zeta_+ - 1) \, \dr_\al R & = & 0,\\
\label{eq:lin_b_0_characSV-}
- \dr_\bt Z + (\zeta_- - 1) \, \dr_\bt R & = & 0,
\end{eqnarray}
\end{subequations}
with the characteristic directions given in the case of $b=0$ by
\begin{equation}
\label{eq:eq:lin_b_0_zeta}
\zeta_+=1, \quad \zeta_-=-R.
\end{equation}
Since one of the characteristic directions is constant, these
equations can be integrated in a way which is similar to the procedure
we used for model with a constant velocity profile in section
\ref{sec:constant_b_zero}. From this calculation one finds easily
that the general solution of equations (\ref{SVUlinearfinal}) with $b=0$
reads
\begin{subequations}
\label{eq:gen_sol_lin_b_0}
\begin{eqnarray}
\label{tofaabbsmallRb=0}
t(\al,\bt) & = & \int_{-\al}^\bt
   \frac{d\bt'}{1+R(\alpha,\beta')},\\
\label{xofaabbsmallRb=0}
x(\al,\bt) & = & -\bt + t(\al,\bt), \\
\label{RofaabbsmallRb=0}
R(\al,\bt) & = & -1 + \sqrt{[R_0(\al)+1]^2 + 2[Z_0(\bt) - Z_0(-\al)]},\\
\label{ZofaabbsmallRb=0}
Z(\al,\bt) & = & Z_0(-\bt).
\end{eqnarray}
\end{subequations}
Studying the configurations we studied in section
\ref{sec:constant_b_zero} for model $\mathcal{P}$ with a constant
velocity profile, we choose $Z_0(x)=mx$ so that the integral in
equation (\ref{tofaabbsmallRb=0}) can be computed explicitly. One obtains, using 
(\ref{RofaabbsmallRb=0}):
\begin{eqnarray} 
\label{eq:lin_b_0_t-R-rel}
t(\al,\bt) & = & \frac{1}{m} \left [ \sqrt{[R_0(\al)+1]^2 +
    2m(\al+\bt)} - R_0(\al)-1
\right ]\nonumber\\
&=& \frac{1}{m}\left[ R(\alpha,\beta)- R_0(\alpha) \right].  
\end{eqnarray}

Since in the limit $b=0$ equation (\ref{SVUlinearfinalZ}) acquires a
simple linear form, we have obviously the result
\begin{equation}
\label{eq:L_solution_Z}
Z(t,x)=Z_0(x-t)=m(x-t)\, .  
\end{equation}
At sufficiently large times one has $\alpha \sim x+mt^2/2$, and
thus equation (\ref{eq:lin_b_0_t-R-rel}) shows that the amount of rolling
grains is given by
\begin{equation}
\label{eq:lin_profile_b_zero_R_at_large_t}
R(t,x)=R_0(\tilde x)+ m t \quad \mbox{with} \quad \tilde x \equiv x+\frac{m}{2} t^2\, .
\end{equation}

\subsubsection{Physical discussion}

From the above result, the shape of the perturbation at a given large
time would be the same for all initial slopes $m$. From this result,
the maximum of the mobile layer $R(t,x)$ travels with a velocity
$v_\text{max}=r_0 + m t/2$ which, for $m >0$, increases linear in
time, i.e., the perturbation feels a {\em constant acceleration}. This
has to be compared to the {\em constant velocity} we found for the
model with a constant velocity profile, see section
\ref{sec:constant_b_zero}. However, for negative $m$, as explained
above, the actual solution is obtained by setting $R(t,x)$ to zero in
regions where it would be negative otherwise. At the time at which
$R(t,x)$ becomes zero, the profile $Z(t,x)$ is frozen at its present
height. Due to this construction the final solution for $Z(t,x)$ will
deviate from the simple form of equation (\ref{eq:L_solution_Z}). For
positive $m$ the profile $R(t,x)$ as obtained from equation
(\ref{eq:lin_b_0_t-R-rel}) is always non-negative.  But there is the
possibility that shocks actually prevent the system to reach the
asymptotic time result of equation
(\ref{eq:lin_profile_b_zero_R_at_large_t}). In fact, independent of
the initial parameters of $R_0$ and $m$, a shock {\em always} occurs
after a finite time $t_s$, unless the solution $R(t,x)$ of equation
(\ref{eq:lin_profile_b_zero_R_at_large_t}) becomes formally negative
(for negative $m$) before the shock can appear. The shock time is
given by the general expression
\begin{equation}
\label{eq:lin_profile_b_zero_shock_time}
t_s= \frac{1}{\text{max}\, R_0'(x)}.
\end{equation}
Remember that for the plug flow $\mathcal P$ model (see
section \ref{sec:constant_profile}), a shock appears only for slopes $m$
which are larger than a positive critical slope. For the model
discussed here, shocks can even occur at negative slopes $m$.

%
\begin{figure}[t] 
\includegraphics[width=0.48\linewidth]{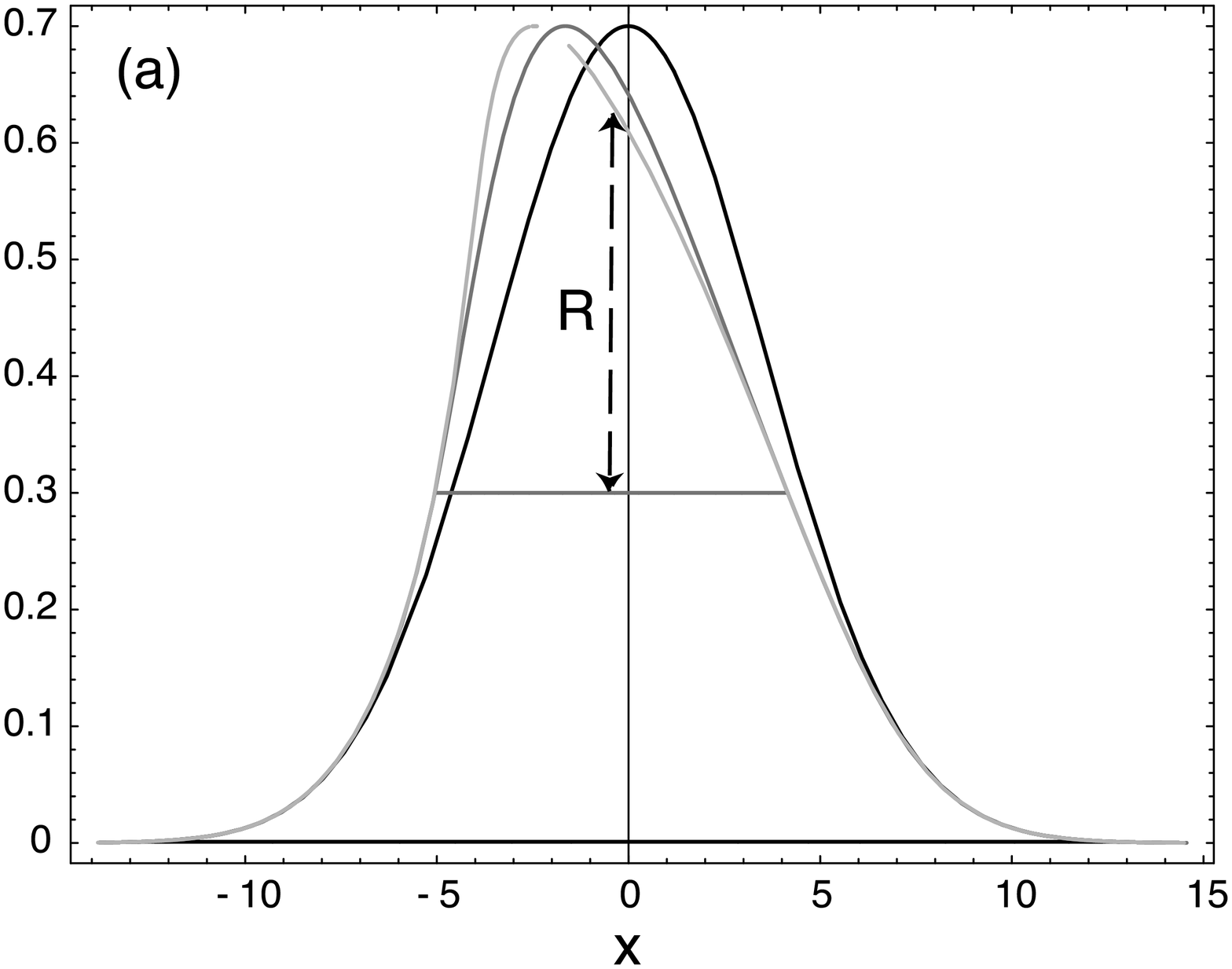}
\hfill
\includegraphics[width=0.48\linewidth]{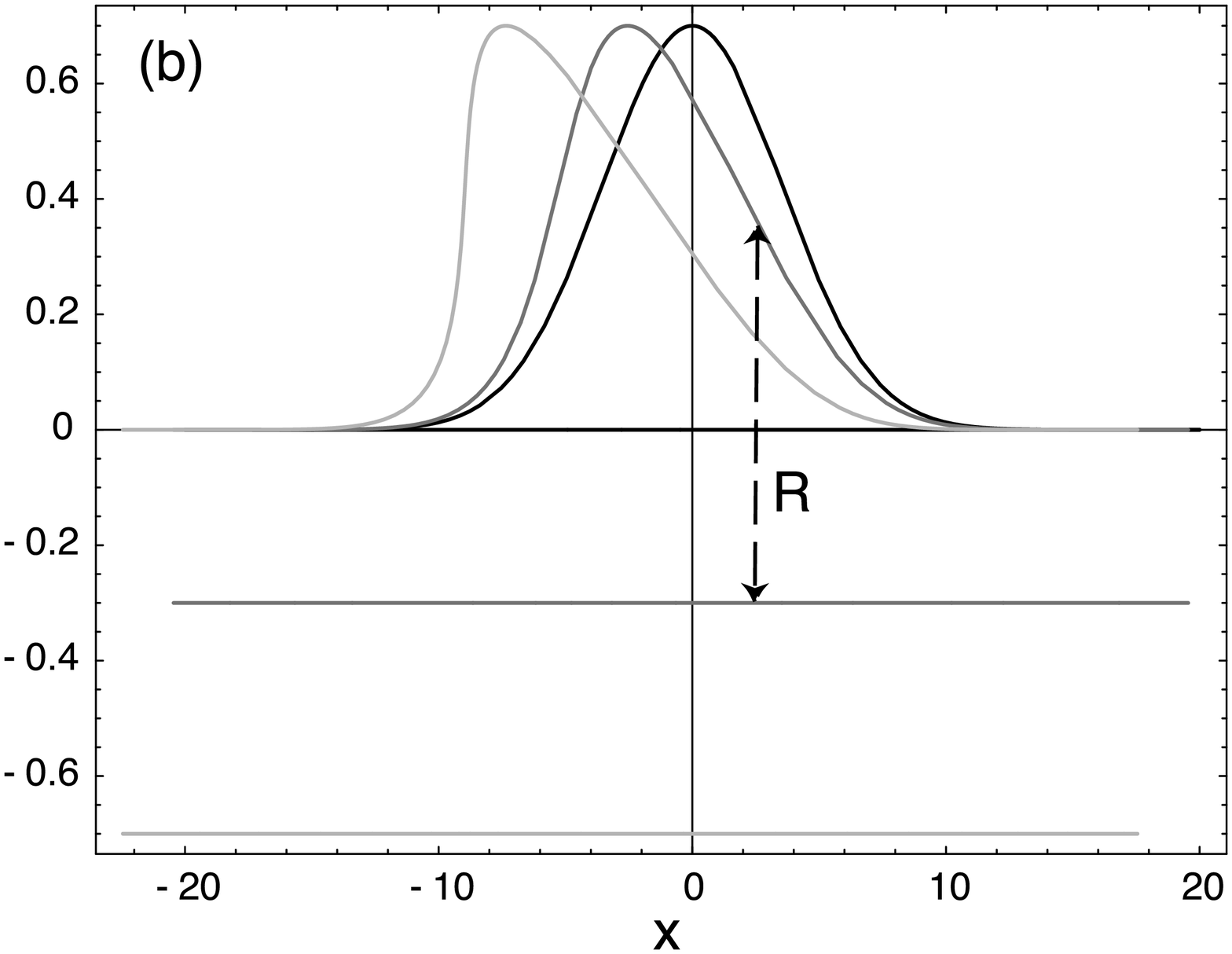}
\caption{Model $\mathcal{L}$ with $b=0$: Profiles for a Gaussian profile $R_0(x)$
  with $r_0=0.7$, $\delta=5.0$ and slope (a) $m=-0.1$ (b) $m=0.1$,
  both for times $t=0$, $3$ and $7$. The shock time is
  $t_s=\sqrt{e/2}\delta/r_0=8.32$. For negative $m$ the perturbation
  has stopped at the time $t=-r_0/m=7.0$, i.e., the profile shown for
  the latter time is the final static profile. Note that for positive
  $m$ the thickness of the moving layer shows an overall linear
  increase proportional to $mt$ since for a Gaussian profile, $R_0(x)$
  is in fact small but {\it finite} everywhere. For clarity, we
  indicated explicitly the thickness $R$ of the mobile layer at $t=3$.}
\label{fig:profiles_linear_b0}
\end{figure}
%

In the following, we will consider again a Gaussian perturbation in
the layer of rolling grains, $R_0(x)=r_0 \exp(-x^2/\delta^2)$. Then
the shock time is $t_s=\sqrt{e/2}\,\delta/r_0$. Figure
\ref{fig:profiles_linear_b0} shows the time evolution of this
perturbation for both positive and negative $m$. Plotted are the
profiles $Z(t,x)-mx$ and $Z(t,x)-mx+R(t,x)$ so that again the gap
between the profiles corresponds to the layer of rolling grains.  For
negative $m$ all moving grains have come to rest at the time
$t=-r_0/m$ which is smaller than the shock time scale $t_s$ for the
parameters used here.  For positive $m$ there is an uniform increase
in the thickness of the layer of rolling grains, see figure
\ref{fig:profiles_linear_b0}(b).  This increase is linear in time,
$\sim mt$, and is unrelated to the amplitude of the local perturbation
$R_0(x)$. This apparently unphysical result can be understood from the
structure of equation (\ref{SVUlinearfinalR}). Even for a strictly
localized initial $R_0(x)$ which is zero outside a finite interval,
there would be an increase $\sim (\dr_x Z) t$ (for a constant slope)
at all positions $x$, not only there where $R_0(x)$ is non-zero. But
since we divided the original equations by $R$ to obtain equation
(\ref{SVUlinearfinal}), $R=0$ is a trivial solution. The latter
solution should be matched with the finite $R$ solution at the front
of the avalanche. However, by definition, at the front the rolling
layer becomes very thin, and a strictly linear velocity profile is
certainly an oversimplification. Thus, with the model of this Section,
the matching of the two solutions at the front is not justified.
Instead, the dynamical equations should be refined as to describe the
physical processes close to an avalanche front and the thin-to-thick
flow crossover (for example along the lines of \cite{BRG98,ARdG99}).
This we leave to a future work.

\subsection{Strong stress anisotropy: small $b$}
\label{sec:linear_b_finite}

\subsubsection{Analytical results}

%
\begin{figure}[t] 
\includegraphics[width=0.48\linewidth]{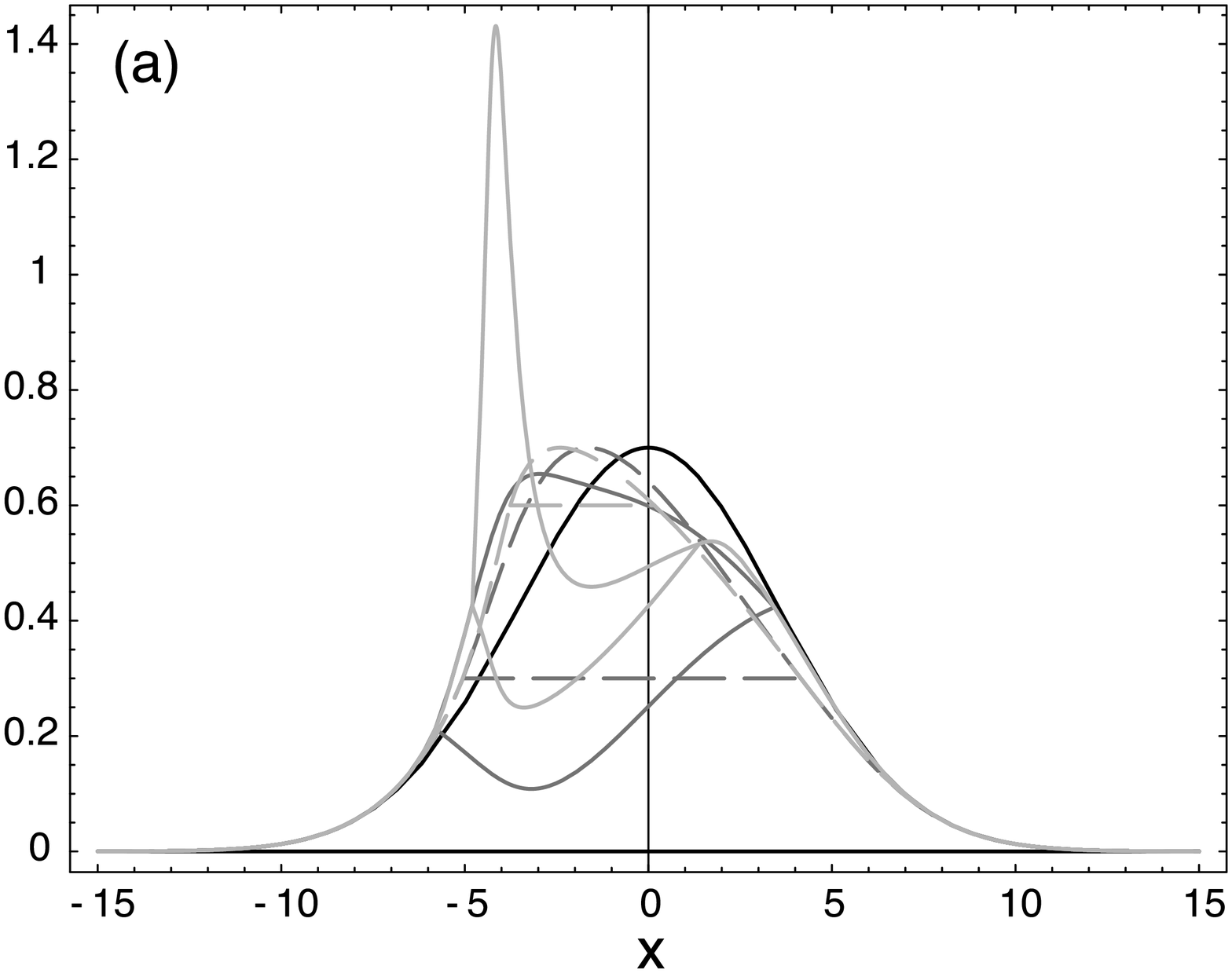}
\hfill
\includegraphics[width=0.48\linewidth]{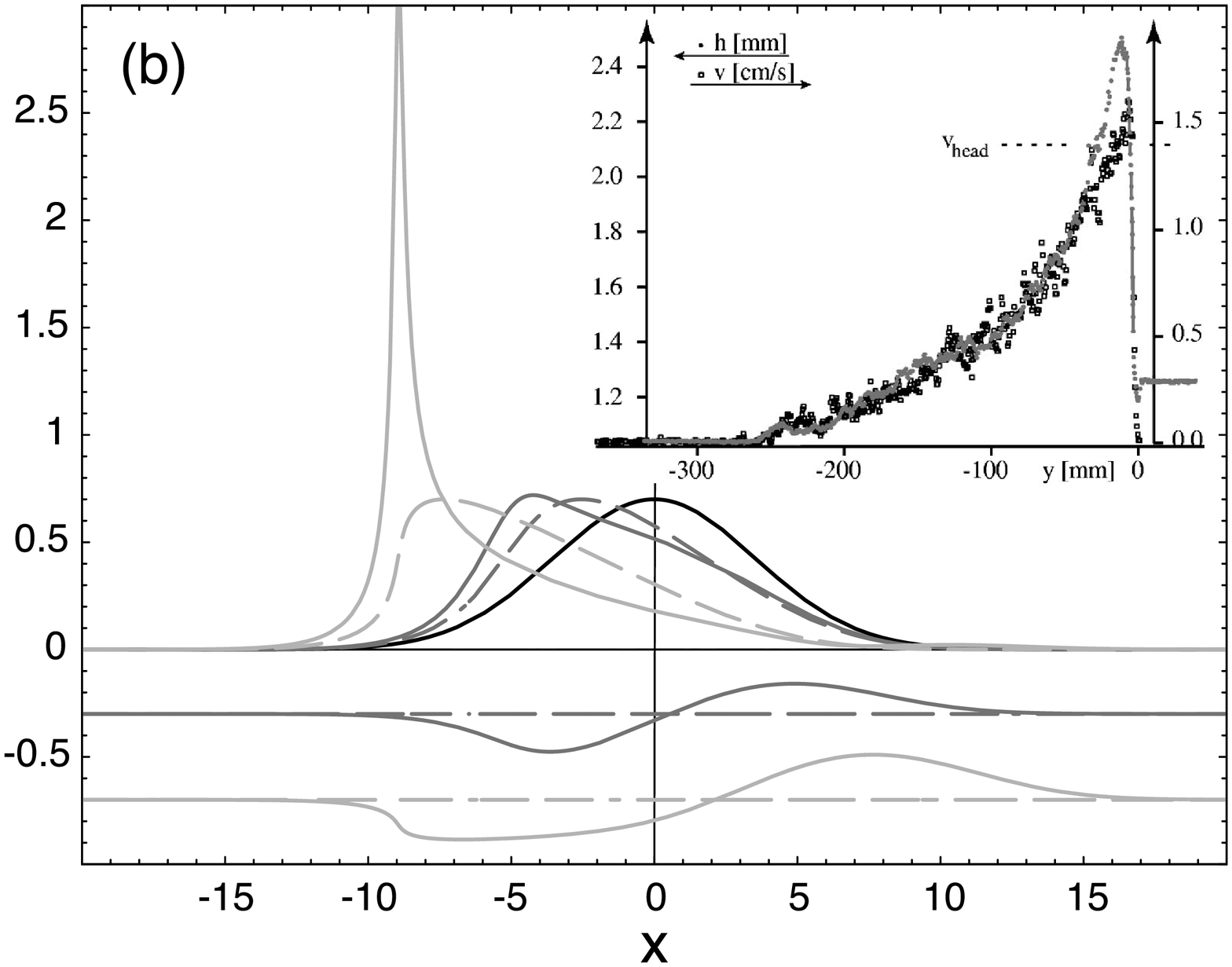}
\caption{Model $\mathcal{L}$ with finite $b=0.5$ for the same parameters as 
  in figure \ref{fig:profiles_linear_b0}. The plots are for times (a)
  $t=0$, $3$ and $6$, and (b) $t=0$, $3$ and $7$. For comparison, the
  corresponding profiles for $b=0$ are shown as dashed curves. Notice
  that the peak appears rather sharp due to the relative elongation of
  the vertical axis. Inset: For qualitative comparison, experimental
  results as given by Fig.~23 of Ref.~\onlinecite{D01} for the surface
  velocity $v$ and the local height $h$ of the mobile layer which
  corresponds to $R$ in our notation. A linear relation between $v$
  and $R$ is observed, and the shape of $R$ resembles that of our
  analytical result.  (The direction of flow is inverted in the
  experiment compared to our model.)}
\label{fig:profiles_linear_b-finite}
\end{figure}
%

Now we study the influence of finite horizontal stress with a finite
but small $b$. An important consequence of a finite $b$ is that now
the equations (\ref{SVUlinearfinal}) become coupled by the stress term. In
order to obtain the dynamic response to a local perturbation we
perturb about the $b=0$ solution of the previous section. Following
the analysis of section \ref{sec:constant_b_finite}, we make the ansatz
\begin{subequations}
\begin{eqnarray}
  \label{eq:linear_ansatz_finite_b}
  Z &=& Z_1 + b Z_2 \\
  R &=& R_1 + b R_2,
\end{eqnarray}
\end{subequations}
where $Z_1$ and $R_1$ denote the solution for $b=0$ of the previous section,
cf equations (\ref{RofaabbsmallRb=0}) and (\ref{ZofaabbsmallRb=0}). By
expansion of the equations (\ref{SVUlinearfinal}) in $b$ we obtain the
dynamics of the contributions from finite $b$,
\begin{subequations}
\begin{eqnarray}
  \label{eq:linear_2nd_order_system}
  \partial_t Z_2 &=& -\partial_x Z_2 - \partial_x R_1 \\
  \partial_t R_2 &=& (1+R_2) \partial_x R_1 + R_1 \partial_x R_2 + 
  \partial_x Z_2.
\end{eqnarray}
\end{subequations}
To this coupled system of equations we can again apply the method of
characteristic curves. The resulting characteristic equations for the
corrections to the profiles are
\begin{subequations}
\label{eq:lin_b_finite_ceqs}
\begin{eqnarray}
\label{eq:lin_b_finite_characSV+}
\dr_\al Z_2 + \dr_x R_1 \, \dr_\al t & = & 0,\\
\label{eq:lin_b_finite_characSV-}
\dr_\bt Z_2 + (1+R_1) \, \dr_\beta R_2 - \left[
(1+R_1) R_2 + R_1 \right] \dr_x R_1 \, \dr_\beta t & = & 0,
\end{eqnarray}
\end{subequations}
where the characteristic direction are the same as in the unperturbed
case, $\zeta_+=1$, $\zeta_-=-R_1$. Again all functions in the above
equations have to be regarded as depending on $\alpha$, $\beta$.
Derivatives with respect to $x$ can be rewritten by the use of the
relations
\begin{equation}
  \label{eq:lin_dr_relations}
 \dr_x \beta = - \left( \frac{R_1}{1+R_1} \right)^2, \quad
 \dr_x \alpha = \frac{\dr_\alpha R_1 - R'_0(\alpha)}{m(1+R_1)}.
\end{equation}
Using the latter relations, equation (\ref{eq:lin_b_finite_characSV+}) can
be integrated with respect to $\alpha$. The result is
\begin{equation}
  \label{eq:lin_final_Z2}
  Z_2(\alpha,\beta)=-\int_{-\beta}^\alpha d\alpha' \frac{R'_0(\alpha')}
{1+R_1(\alpha',\beta)}.
\end{equation}
By inserting this solution into equation (\ref{eq:lin_b_finite_characSV-})
one obtains an ordinary differential equation (with respect to
$\beta$) for $R_2(\alpha,\beta)$. In formal analogy to
equation (\ref{eq:bcre_result_Z_2}) its solution can be written as
\begin{equation}
\label{eq:lin_integrals_R2}
  R_2(\alpha,\beta) = 
\exp\left[ \int_{-\alpha}^\beta d\beta' g_1(\alpha,\beta') \right]
\int_{-\alpha}^\beta d\beta' g_2(\alpha,\beta') \exp\left[
-\int_{-\alpha}^{\beta'} d\beta'' g_1(\alpha,\beta'') \right],
\end{equation}
where the functions $g_1$ and $g_2$ are now given by
\begin{subequations}
  \begin{eqnarray}
    \label{eq:lin_g+h}
    g_1(\alpha,\beta) &=& \frac{m}{(1+R_1)\left[ R_0(\alpha) + 
      m/R'_0(\alpha)-R_1\right]} \\
    g_2(\alpha,\beta) &=& \frac{1}{1+R_1} \left[ R_1 g(\alpha,\beta) -
    \dr_\beta \ln(1+R_0(-\beta)) +
    m \int_\alpha^{-\beta} d\alpha' 
    \frac{R'_0(\alpha')}{(1+R_1(\alpha',\beta))^3}\right] \, .
  \end{eqnarray}
\end{subequations}
Most of the integrals in equation (\ref{eq:lin_integrals_R2}) can be
obtained in closed form. After a tedious calculation the final result
can be expressed in terms of single integrals, that we give for completeness:
\begin{eqnarray}
  \label{eq:lin_final_R2}
  &R_2(\alpha,\beta)&=\frac{1}{1+(R_0(\alpha)-R_1)R'_0(\alpha)/m}
\left\{ \frac{R'_0(\alpha)}{m} \left[ R_1 - R_0(\alpha) -\ln 
\left( \frac{(1+R_1)(1+R_0(-\beta))}{(1+R_0(\alpha))^2}\right)\right]\right.
\nonumber\\
& + &  \!\!\!\!\!\!\!\!\! \left. \int_{-\alpha}^\beta d\beta' \frac{R'_0(-\beta')
(1+(R_0(\alpha)-
R_1(\alpha,\beta'))R'_0(\alpha)/m}{(1+R_1(\alpha,\beta'))(1+R_0(-\beta'))}
-\frac{R'_0(\alpha)}{m}\int_{-\beta}^\alpha d\alpha' 
\frac{R'_0(\alpha')}{1+R_1(\alpha',\beta)}\right. \nonumber \\
& - & \!\!\!\!\!\!\!\!\! \left. \left[ 1+(R_0(\alpha)+1)R'_0(\alpha)/m \right]
\int_{-\beta}^\alpha  \frac{R'_0(\alpha')d\alpha'}{(R_1(\alpha,\beta)+1)^2
-(R_1(\alpha',\beta)+1)^2} \left[ \frac{1+R_1(\alpha,\beta)}
{1+R_1(\alpha',\beta)} - \frac{1+R_1(\alpha,-\alpha')}{1+R_0(\alpha')} \right]
\right\} \,\,
\end{eqnarray}
The equations (\ref{eq:lin_final_Z2}) and (\ref{eq:lin_final_R2}) provide
the final result for the changes in the profiles due to a finite
horizontal stress. The solution is valid for \emph{general} initial
profiles $R_0(x)$. In figure \ref{fig:profiles_linear_b-finite} we have
plotted this solution for a Gaussian perturbation with the same
parameters as in the previous section, cf figure \ref{fig:profiles_linear_b0}.

\subsubsection{Physical discussion}

The main difference coming from finite $b$ is the generation of a peak
at the downhill front of the avalanche, both for positive and negative
slopes $m$.  From the model of Eqs.~(\ref{SVUlinearfinal}) one
observes that the term proportional to $b$ is controlled by the slope
of the moving layer. Since this slope becomes steeper at the downhill
front with evolving time, there is an increasing thickness of the
mobile layer close to the front. From a physical point of view this
amplification can be understood from the effect of the horizontal
stress which increases with $R$. Thus at the downhill front there is a
net force which pushes the material towards the front and induces an extra
growth of the rolling layer. As demonstrated by the inset of
Fig.~\ref{fig:profiles_linear_b0} our result is in qualitative
agreement with measurements of the thickness of the mobile layer of
grains along the symmetry axis of a triangular shaped avalanche moving
on a static layer of limited thickness \cite{D01}.  For positive $m$
there is, in addition to the homogeneous and linear decrease of $Z$, a
net transport of grains of the sand bed from the downhill front to the
uphill end.

\section{Conclusion and discussion}
\label{sec:discussion}

In this paper, we have studied two sets of St Venant equations for the
modeling of granular avalanches on an erodible bed. The models differ
in the choice of the velocity profile within the flowing layer --
either a plug flow ($\mathcal{P}$) or a profile with a constant
velocity gradient ($\mathcal{L}$) -- which give rise to different
non-linearities. These models can been solved analytically by the
method of characteristics, at least for sufficiently large stress
anisotropy, and we have focused our attention on the situation where a
uniform static slope is initially disturbed by a localized amount of
rolling grains. We were able to compute the space and time evolution
of the avalanche which either dies or grows, depending on whether the
initial slope is below or above the angle of repose.

Such unsteady and non-uniform flows are very demanding for the models
as the description of fronts and the generation of shocks must be
addressed. For the case of a plug flow we found that shocks occur at
the uphill end of the avalanche above a positive critical value $m_c$
of the slope. Below this value, the asymptotic large time behavior can
be computed, and we found that the amplitude and width of the
avalanche grow linearly in time if $m>0$, whereas for negative $m$ the
initial perturbation decays exponentially and the width of the
deposited bump of grains scales like $1/m$. For a linear velocity
profile, shocks occur for all slopes at a finite time that can be
computed explicitly and is related to the shape of the initial
distribution of rolling grains. By contrast to the previous case,
shocks are located at the front of the avalanche.

Our analysis shows that these models predict several interesting
qualitative features of granular avalanches, that can be compared with
experiments.  However, in the present form they also certainly have a
number of shortcoming. The plug flow assumption, for example, is not
tenable for a rolling layer thickness which starts to be of the order
of several grain diameters, as particles on the top of such a layer
would not feel the damping due to the friction on the static bed. On
the other hand, the linear velocity hypothesis yields a vanishing
velocity $U \to 0$ when the thickness of the mobile layer $R \to 0$,
which forbids any front to move. However, shock-less and well defined
propagative fronts are observed experimentally in steady \cite{P99} or
unsteady \cite{DACD01} situations. Our model also does not correctly
account for slope hysteresis which should differentiate between a
starting and a stopping angle, $\phi_{\mbox{\scriptsize start}}$ and
$\phi_{\mbox{\scriptsize stop}}$, respectively. In fact, the region
between $\phi_{\mbox{\scriptsize stop}}$ and $\phi_{\mbox{\scriptsize
    start}}$ is precisely that of major interest as static slopes with
an angle $\phi > \phi_{\mbox{\scriptsize stop}}$ are stable but can
generate growing avalanches when disturbed by a local amount of
rolling grains. Because the non-linear terms in model $\mathcal{P}$
are all proportional to $R$, spontaneous avalanches (for $R=0$) can
never occur whatever the value of the initial slope.  This would
correspond to the maximal value of $\pi/2$ for
$\phi_{\mbox{\scriptsize start}}$.  This, however, implies that model
$\mathcal{P}$ can be applied to the experimentally important regime of
slopes which are only slightly larger than $\phi{\mbox{\scriptsize
    stop}}$ and sufficiently small compared to
$\phi_{\mbox{\scriptsize start}}$.  On the other hand, the uniformly
growing solution for $R$ in model $\mathcal{L}$ can be interpreted as
evidence that both angles $\phi_{\mbox{\scriptsize start}}$ and
$\phi_{\mbox{\scriptsize stop}}$ are identical so that beyond that
angle avalanches occur even at $R=0$ (unstable slope).

To include the above mentioned effects in the presented models, new
directions have to be proposed. First, one can modify the dynamical
equations itself while keeping their general structure (hyperbolic
first order differential equations) with its analytic properties being
tractable by the method of characteristics. Second, additional
physical input could be used to either study the propagation of shocks
beyond the shock time scale or one could implement boundary conditions
on, e.g., the slope of the profiles to describe the dynamics close to
the avalanche front. To be more specific, we discuss two points of
particular interest: The velocity profile and hysteresis effects.

As proposed theoretically and demonstrated by experiments, an
$R$-dependence of the flow velocity must be kept to allow for thick
avalanches. The commonly used strictly linear velocity profile with
the rheological ansatz $u(x,y)=\gamma y$ (constant shear rate) leads
to problems at the avalanche front since the depth averaged velocity
$U$ should stay finite and of the order of $\sqrt{gd}$ when $R \to 0$.
A possibility is to let $\gamma$ diverge in this limit. This would
correspond to a crossover from model $\mathcal{L}$ to model
$\mathcal{P}$ below a certain small value for $R$ of the order of a
grain diameter $d$ (see \cite{ARdG99}). The corresponding matching of
characteristic curves is technically involved and moreover such a
treatment would not provide an understanding of the physical mechanism
of velocity selection.  In fact, close photos of the foot of the
avalanche fronts reported in \cite{P99} show a small gas-like region
where grains are ejected from the dense flow.  This means that this
zone is, in a sense, outside of the present modeling framework and thus
could allow for a discontinuity in $R$, for example, or for imposing
an extra constraint on the profiles $R$ and $Z$ or their derivatives
-- e.g. a fixed slope of the free surface as observed on propagative
fronts. Another more fundamental approach would be to consider the
velocity $U$ as an independent dynamical field which a priori is not
related to $R$ by a fixed velocity profile. Such kind of description
has been applied to granular flow on a {\it fixed} plane \cite{PF02}
so that $Z$ is not a dynamical quantity. 

Another feature of sand piles that should be included in the St Venant
models discussed here is the hysteresis of avalanche dynamics as
reflected by the existence of two critical angles. Experimentally it
is observed that, at least for a fixed profile $Z$, the critical
angles are actually functions of the thickness $R$ of the mobile layer
\cite{PF02}. Since the friction coefficient $\mu$ is given by the
tangent of the actual angle of the pile, a non-constant friction
coefficient $\mu(R)$ is expected.  Indeed, at the scale of a single
grain, the starting angle depends on the depth of ``traps'' due to the
roughness of the static bed \cite{QADD00}.  Therefore one should
expect an increased value of $\mu$ below a typical thickness
$R_{\mbox{\scriptsize trap}}$ (of the order of a grain diameter).
This increased value determines then the staring angle
$\phi_{\mbox{\scriptsize start}}$ while the $\mu$ at large $R$
corresponds to the stopping angle $\phi_{\mbox{\scriptsize stop}}$.  A
sufficiently large value of the friction coefficient at small $R$
would lead, for the points where $R \le R_{\mbox{\scriptsize trap}}$,
to a ``freezing'' of the sand bed profile $Z$ at its current value.

Both modifications, constraints on the profiles at the avalanche front
and a friction function $\mu(R)$, do not change the general structure
of the dynamical equations studied in the present paper. Thus, the
method of characteristics and our predictions should prove useful for
a better understanding and modeling of non-stationary granular flow.

\section*{Acknowledgments}

We thank B. Andreotti for essential discussions and a careful reading
of the manuscript.  This work was supported by the PROCOPE program of
EGIDE (P.C. and J.-P.B.) and DAAD (T.E.) and by the Deutsche
Forschungsgemeinschaft through the Emmy Noether grant No. EM70/2-2
(T.E.). The LPMMH is UMR 7636 of the CNRS.  P.C. was before
associated with the Laboratoire des Milieux D\'esordonn\'es et
H\'et\'erog\`enes (UMR 7603).


\begin{appendix}

\section{Method of characteristic curves}
\label{sec:app_cc}

In this section a brief account of the general theory for systems of
partial differential of first order and {\it hyperbolic} type is
presented. Due to the relevance to granular flow problems we will
concentrate on non-linear systems consisting of two equations for two
functions of two independent variables. In the present context of this
paper, the functions are $R$ and $Z$ and the independent variables
correspond to space $x$ and time $t$. For such systems a complete
mathematical theory is available \cite{Courant-Friedrichs-book}. Our
presentation will follow closely the latter reference. 

For hyperbolic systems the notion of characteristic curves is the
central concept. Before introducing the general theory, we would like
to motivate the introduction of characteristic curves or coordinates.
This concept is particularly adapted to the case where the number of
equations equals the number of independent variables. In the present
case of two equations, the objective of the method of characteristics
is to introduce instead of $(t,x)$ a new coordinate frame
$(\alpha,\beta)$ so that along the two families of curves of constant
coordinates $\alpha$ and $\beta$ the partial differential equations
reduce to {\it ordinary} differential equations with respect to
$\alpha$ and $\beta$.

Let us demonstrate the method explicitly for the simple case of one
linear partial differential equation for a function $f(t,x)$ of the
form
\begin{equation}
  \label{eq:cc-simple-example}
  a(t,x) \, \dr_x f + b(t,x) \, \dr_t f + c(t,x) \, f = 0.
\end{equation}
The initial value will be prescribed at zero time, $f(t=0,x)=f_0(x)$.
First, we have to define the characteristic curves
$[t(\alpha),x(\alpha)]$ where $\alpha$ is a variable parameterizing a
given curve. These curves are specified in terms of their local
tangent vectors (velocities),
\begin{equation}
  \label{eq:cc-simple-def}
  \frac{dx}{d\alpha} = a(t,x), \quad \frac{dt}{d\alpha} = b(t,x)\, .
\end{equation}
Integrating this equations yields a whole family of characteristic
curves which is parameterized by the starting position $x_0$ of the
curves at $t=0$ so that $x(\alpha=0)=x_0$ and $t(\alpha=0)=0$. This
ensures that from each position of the line $t=0$ exactly one
characteristic curve originates, providing the trajectory along which
the initial data $f_0(x)$ can be propagated in time.  In order to see
why the definition of Eq.~(\ref{eq:cc-simple-def}) is useful we
compute the change of $f$ along the characteristic curves, yielding
\begin{equation}
  \label{eq:cc-change-of-f}
  \frac{df}{d\alpha} + c(t,x)\, f = 0\, .
\end{equation}
The crucial observation is that the latter equation is just an {\it
  ordinary} differential equation, which is valid along the
characteristic curves. To find the final solution $f(t,x)$ to
Eq.~(\ref{eq:cc-simple-example}) one proceeds as follows. First, one
solves Eq.~(\ref{eq:cc-simple-def}) to obtain the relation between
$(t,x)$ and $(\alpha,x_0)$. Second, the ordinary differential
Eq.~(\ref{eq:cc-change-of-f}) is solved with the initial condition
$f(\alpha=0)=f_0(x_0)$ which provides the solution $f(\alpha,x_0)$.
Finally, the parameters $\alpha$ and $x_0$ are computed for a given
coordinate $(t,x)$ to get the solution $f(t,x)$ in the original
coordinate frame.

Having outlined the general idea behind the method of characteristics,
we can go ahead and turn to the case of two non-linear hyperbolic
equations for two functions. We will consider a general system of the
form
\begin{subequations}
  \label{eq:cc-general-system}
  \begin{eqnarray}
    L_1 & = & A_1 \, \dr_x R + B_1 \, \dr_t R + C_1 \, \dr_x Z + 
    D_1 \, \dr_t Z + E_1 = 0\\
    L_2 & = & A_2 \, \dr_x R + B_2 \, \dr_t R + C_2 \, \dr_x Z + 
    D_2 \, \dr_t Z + E_2 = 0
  \end{eqnarray}
\end{subequations}
for the functions $R(t,x)$ and $Z(t,x)$, where the coefficients $A_1$,
$A_2$, $B_1$, $\ldots$, are known functions of $x$, $t$, $R$ and $Z$.
The type of this system depends on the coefficients. For the
hyperbolic case in which we are interested here one needs that $ac-b^2
< 0$ with the functions
\begin{equation}
  \label{eq:cc-def-abc}
  a = [A,C], \quad 2b = [A,D]+[B,C], \quad c = [B,D]\, ,
\end{equation}
where $[X,Y]=X_1 Y_2 - X_2 Y_1$. 

The goal is again to reduce the above system to a system of ordinary
differential equations with respect to new coordinates $\alpha$,
$\beta$. Since we have now to deal with two unknown functions $R(t,x)$
and $Z(t,x)$ we start by searching for a linear combination
$L=\lambda_1 L_1 +\lambda_2 L_2$ of the differential operators in
Eq.~(\ref{eq:cc-general-system}) so that the derivatives of $R$ and
those of $Z$ combine to derivatives in the same direction. These
directions will be the velocity vectors of the characteristic curves
and thus determine the new coordinate frame. Let us represent an
arbitrary curve in the $x-t$ plane by $[t(\sigma),x(\sigma)]$ with
$\sigma$ denoting the running parameter along the curve --- note that
$\sigma$ finally will play the role of $\alpha$ or $\beta$. Then the
condition that in $L$ both functions $R$ and $Z$ are differentiated in
the tangential direction of this curve reads
\begin{equation}
  \label{eq:cc-same-direct-cond}
  \frac{dx/d\sigma}{dt/d\sigma} = 
  \frac{\lambda_1 A_1+\lambda_2 A_2}{\lambda_1 B_1 + \lambda_2 B_2} = 
  \frac{\lambda_1 C_1+\lambda_2 C_2}{\lambda_1 D_1 + \lambda_2 D_2} \, .
\end{equation}
Next, we consider the change of the functions $R$ and $Z$ along the
curve $[t(\sigma),x(\sigma)]$. It is given by $dR/d\sigma = \dr_x R \,
dx/d\sigma + \dr_t R \, dt/d\sigma$ and analogous for $Z$. Multiplying
$L$ with either $dx/d\sigma$ or $dt/d\sigma$ and using the conditions of
Eq.~(\ref{eq:cc-same-direct-cond}) one gets
\begin{subequations}
  \label{eq:cc-chnage-of-fcts-along-curve}
  \begin{eqnarray}
    \frac{dx}{d\sigma} L & = & (\lambda_1 A_1 + \lambda_2 A_2 ) 
    \, \frac{dR}{d\sigma} + (\lambda_1 C_1 + \lambda_2 C_2 ) \, 
    \frac{dZ}{d\sigma}
    + (\lambda_1 E_1 + \lambda_2 E_2 ) \, \frac{dx}{d\sigma} \\
    \frac{dt}{d\sigma} L & = & (\lambda_1 B_1 + \lambda_2 B_2 ) \,
    \frac{dR}{d\sigma} + (\lambda_1 D_1 + \lambda_2 D_2 ) \, \frac{dZ}{d\sigma}
    + (\lambda_1 E_1 + \lambda_2 E_2 ) \, \frac{dt}{d\sigma} \, .
  \end{eqnarray}
\end{subequations}
If the functions $R$ and $Z$ satisfy the system of differential
Eqs.~(\ref{eq:cc-general-system}) we have $L=0$, and we obtain the
following four homogeneous linear equations for the coefficients
$\lambda_1$ and $\lambda_2$ which result from
Eq.~(\ref{eq:cc-same-direct-cond}) and
Eq.~(\ref{eq:cc-chnage-of-fcts-along-curve}),
\begin{subequations}
  \label{eq:cc-system-for-lambdas}
  \begin{eqnarray}
  \label{eq:cc-system-for-lambdas-1}
    \lambda_1 \left( A_1 \frac{dt}{d\sigma} - B_1 \frac{dx}{d\sigma} \right)
     + \lambda_2 \left( A_2 \frac{dt}{d\sigma} - B_2 \frac{dx}{d\sigma} \right)
        & = & 0 \\
        \label{eq:cc-system-for-lambdas-2}
      \lambda_1 \left( C_1 \frac{dt}{d\sigma} - D_1 \frac{dx}{d\sigma} \right)
     + \lambda_2 \left( C_2 \frac{dt}{d\sigma} - D_2 \frac{dx}{d\sigma} \right)
        & = & 0 \\
        \label{eq:cc-system-for-lambdas-3}
     \lambda_1 \left( A_1 \frac{dR}{d\sigma} + C_1 \frac{dZ}{d\sigma} + 
       E_1 \frac{dx}{d\sigma} \right) + \lambda_2 
     \left( A_2 \frac{dR}{d\sigma} + C_2 \frac{dZ}{d\sigma} + 
       E_2 \frac{dx}{d\sigma} \right) & = & 0 \\
     \label{eq:cc-system-for-lambdas-4}
     \lambda_1 \left( B_1 \frac{dR}{d\sigma} + D_1 \frac{dZ}{d\sigma} + 
       E_1 \frac{dt}{d\sigma} \right) + \lambda_2 
     \left( B_2 \frac{dR}{d\sigma} + D_2 \frac{dZ}{d\sigma} + 
       E_2 \frac{dt}{d\sigma} \right) & = & 0 \, .
  \end{eqnarray}
\end{subequations}
This system is obviously over-determined. Thus, in order to have a
non-trivial solution, the determinant of every pair of rows in the
matrix of coefficients of $\lambda_1$ and $\lambda_2$ has to vanish.
The relations following from this conditions are called {\it
  characteristic relations}. 

In particular, from the first two
Eqs.~(\ref{eq:cc-system-for-lambdas-1}),
(\ref{eq:cc-system-for-lambdas-2}), one obtains the condition
\begin{equation}
  \label{eq:cc-c-rel-1}
  a \left( \frac{dt}{d\sigma} \right)^2 - 2b \, \frac{dx}{d\sigma} 
\frac{dt}{d\sigma} 
+ c \left( \frac{dx}{d\sigma} \right)^2 = 0
\end{equation}
with the coefficients given by Eq.~(\ref{eq:cc-def-abc}). From this
condition it becomes clear why the method of characteristics applies
to hyperbolic systems. For those systems we have, as mentioned above,
$ac-b^2 < 0$, and Eq.(\ref{eq:cc-c-rel-1}) has two different solutions
and thus two different characteristic directions $(dx/d\sigma,
dt/d\sigma)$ through each point. In the following we assume, without
any restrictions, that $a \neq 0$ so that $dx/d\sigma \neq 0$ and we
can introduce the slope
\begin{equation}
  \label{eq:cc-def-slope}
  \zeta=\frac{dt/d\sigma}{dx/d\sigma} \, .
\end{equation}
The two different real solutions of $a\zeta^2-2b\zeta+c=0$ for these
so-called characteristic directions will be denoted by $\zeta_+$ and
$\zeta_-$, respectively. These characteristic directions are in
general functions of $t$, $x$, $R$ and $Z$. Two one-parameter families
of characteristic curves follow from the directions by integration of
the ordinary differential equations $dt/dx = \zeta_+(x,t,R,Z)$ and
$dt/dx = \zeta_-(x,t,R,Z)$. In the following we will denote the
families of curves by $C_+$ and $C_-$. These two families of curves
define a curved coordinate net if the curves are represented as
$\alpha(x,t)=$ constant and $\beta(x,t)=$ constant for family $C_-$
and $C_+$, respectively. The functions $\alpha(x,t)$ and $\beta(x,t)$
are called {\it characteristic parameters}. The coordinates $(t,x)$
corresponding to a given pair $(\alpha,\beta)$ can be obtained as
follows. Consider a curve ${\cal I}$ given by $[x(s),t(s)]$ which has
nowhere a characteristic direction as tangential vector. In practice,
${\cal I}$ will be usually the line $t=0$ where the initial data are
defined. In addition boundary conditions, e.g., a fixed flux at a
given position, can be specified by a curve with $x=$constant.
Through the two points $s=\alpha$ and $s=\beta$ on the curve ${\cal
  I}$ one follows the characteristic curve of family $C_-$ and $C_+$,
respectively, up to the point where the two curves intersect.  The new
coordinates of this intersection point $(t,x)$ are then
$(\alpha,\beta)$. The characteristic parameters $\alpha$ and $\beta$
can now replace the parameter $\sigma$ for the curves of family $C_+$
and $C_-$, respectively, so that one has $dt/d\alpha = \zeta_+
dx/d\alpha$ and $dt/d\beta = \zeta_- dx/d\beta$.

Next, we have to find equations which determine the evolution of the
functions $R$ and $Z$ along the characteristic curves. This can be
done by eliminating $\lambda_1$ and $\lambda_2$ from the
Eqs.(\ref{eq:cc-system-for-lambdas-1}) and
(\ref{eq:cc-system-for-lambdas-3}). Using $dt/d\sigma = \zeta
dx/d\sigma$, where $\zeta$ denotes either $\zeta_+$ or $\zeta_-$ and
$\sigma$ is either $\alpha$ or $\beta$, one obtains
\begin{equation}
  \label{eq:cc-relation-for-R+Z}
  T \frac{dR}{d\sigma} + (a\zeta-S) \frac{dZ}{d\sigma} +
  (K \zeta -H) \frac{dx}{d\sigma} = 0 \, ,
\end{equation}
with the coefficients
\begin{equation}
  \label{eq:cc-coeff-2}
  T=[A,B], \quad S=[B,C], \quad K=[A,E], \quad H=[B,E] \, .
\end{equation}
If we apply the latter equation to the curves of $C_+$ and $C_-$ and
combine them with the equations for the characteristic curves, we
finally obtain the following four {\it characteristic equations} which
are differential equations for the four functions $x(\alpha,\beta)$,
$t(\alpha,\beta)$, $R(\alpha,\beta)$ and $Z(\alpha,\beta)$ and replace
the original system of Eq.~(\ref{eq:cc-general-system}),
\begin{subequations}
  \label{eq:cc-charac-eqs}
  \begin{eqnarray}
    \dr_\alpha t - \zeta_+ \dr_\alpha x & = & 0\, , \\
    \dr_\beta t -\zeta_- \dr_\beta x & = & 0\, , \\
    T \dr_\alpha R + (a\zeta_+ - S)\dr_\alpha Z +(K\zeta_+ - H) 
    \dr_\alpha x & =& 0 \, , \\
    T \dr_\beta R + (a\zeta_- - S)\dr_\beta Z +(K\zeta_- - H) 
    \dr_\beta x & =& 0 \, .
  \end{eqnarray}
\end{subequations}
All the coefficients in this system are known functions of $x$, $t$,
$R$ and $Z$. It can be shown that every solution of this
characteristic systems satisfies the original system of
Eq.~(\ref{eq:cc-general-system}) provided that $\dr_\alpha x \dr_\beta
t -\dr_\beta x \dr_\alpha t = (\zeta_- - \zeta_+) \dr_\alpha x
\dr_\beta x$ is non-zero.  With the derivation of
Eq.~(\ref{eq:cc-charac-eqs}) we reached our initial objective to
reduce the partial differential equations to a form which resembles
that of ordinary differential equations along certain curves. This can
be seen from the fact that each equation contains derivatives with
respect to only one of the coordinates $\alpha$ and $\beta$. Moreover,
the system has to the convenient property that the coefficients do not
dependent on the independent variables $\alpha$ and $\beta$.

Now we are in the position to outline the strategy for solving an
initial value problem for the system of
Eq.~(\ref{eq:cc-general-system}). Let us assume that the initial
values of the functions $R$ and $Z$ are given on the line $t=0$ by
$R_0(x)$ and $Z_0(x)$, and that this line has no characteristic
directions. This line corresponds then to the curve ${\cal I}$
introduced above. We may consider this curve as the image of the
characteristic parameters obeying the relation $\alpha+\beta=0$. Then
we have to solve the system of Eq.~(\ref{eq:cc-charac-eqs}) with the
initial conditions
\begin{equation}
  \label{eq:cc-initial-cond}
  t(\alpha,-\alpha) = 0, \quad x(\alpha,-\alpha) = \alpha, \quad
  R(\alpha,-\alpha) = R_0(\alpha), \quad Z(\alpha,-\alpha) = Z_0(\alpha) \, .
\end{equation}
Due to the particular simple structure of the system of
Eq.(\ref{eq:cc-charac-eqs}), this problem can be treated as completely
as the initial value problem for ordinary differential equations. It
is this formulation of the non-linear hyperbolic flow problem which we
used throughout the paper to solve the partial differential equations
exactly. Finally, we note that this method can be generalized to an
arbitrary number $n$ of equations. Then the system will have $n$
characteristic directions $\zeta_n$ and correspondingly $n$ different
families of characteristic curves. However, the $n$ resulting
characteristic parameters can no longer be interpreted as a new
coordinate frame since there are only the two coordinates $t$ and $x$.

\section{Derivation of the shock condition}
\label{sec:app_shock}

In this appendix we first review the mechanism for the generation of
shocks and their mathematical definition. Then we provide for model
${\cal P}$ the details of the calculations for the shock existence
criterion and the time and position of the shock. For model ${\cal L}$
shocks are always generated, and we derive the simple result of
Eq.~(\ref{eq:lin_profile_b_zero_shock_time}) for the shock time. In
Appendix ~\ref{sec:app_cc} we have assumed that the characteristic curves
of one family (either $C_+$ or $C_-$) do not intersect. Only if this
is true there is a well defined mapping between the original
coordinates $(t,x)$ and the characteristic parameters
$(\alpha,\beta)$. However, depending on the initial data at zero time,
it is possible that characteristics of the same family intersect at a
finite time. Beyond this shock time the system of partial differential
equations fails to have a single valued solution but only multi-valued
solutions or even no solution at all exists at later times. The points
of intersection of characteristic curves are enclosed by an envelope,
cf.~Fig.\ref{fig:c-curves}. The earliest time where a shock appears is
the position of the cusp of this envelope. Technically, the envelope
is defined by the condition that for every position on the envelope
there exists a characteristic curve that touches the envelope at the
position so that both curves have the same tangential direction. If we
represent the envelope as $[t_e(\alpha),x_e(\alpha)]$ where $\alpha$
is used as the parameter changing along the envelope then the
conditions read
\begin{equation}
  \label{eq:shock-env-def}
  x_\alpha(t_e(\alpha))=x_e(\alpha), \quad 
  \partial_\alpha x_\alpha(t)|_{t=t_e(\alpha)} =0.
\end{equation}
where $x_\alpha(t)$ is the trajectory of the characteristic curve
along which $\alpha$ is constant. The second condition follows from
the requirement that the tangent vector $[1,\dr_t x_\alpha(t)]$ of the
curve $x_\alpha(t)$ is parallel to the tangent
$[dt_e(\alpha)/d\alpha,dx_e(\alpha)/d\alpha]$ of the envelope. To see
this, one takes the derivative of the first condition of
Eq.~(\ref{eq:shock-env-def}) with respect to $\alpha$ so that
one obtains
\begin{equation}
  \label{eq:shock-env-cond}
  \frac{dx_e(\alpha)}{d\alpha} = \dr_t x_\alpha(t) \frac{dt_e(\alpha)}{d\alpha}
  + \dr_\alpha x_\alpha(t)|_{t=t_e(\alpha)} \, .
\end{equation}
The collinearity of the two tangent vectors requires then the last
term on the rhs of Eq.~(\ref{eq:shock-env-cond}) to vanish.  Notice
that in model ${\cal P}$ the characteristic curves along which $\beta$
is constant are straight lines, and thus they can never produce a
shock.

In the following we focus on the initial profile $R_0(x)$ given by
Eq.~(\ref{eq:BCRE-R0}) which allows for an explicit calculation of the
envelope and the condition for shocks. For this profile it is useful
to consider three different sectors in the $t-x$ plane, see
Fig.~\ref{fig:sectors}.  Using the above conditions, one obtains for
the envelope in sector (II) the result
\begin{subequations}
  \label{eq:env-II}
  \begin{eqnarray}
  t_e(\alpha)&=&\frac{1}{m-1/\delta} \ln \left[
-\frac{m}{h'(\alpha)}-\frac{1}{h(\alpha)}\right]\\
x_e(\alpha)&=&\frac{1}{m-1/\delta}\left[
h(\alpha) e^{(m-1/\delta)t_e(\alpha)}+\ln\left(
\frac{h(\alpha)}{r_0}\right) + m \alpha - r_0  
\right]\, ,
\end{eqnarray}
\end{subequations}
where the function $h(\alpha)$ is given by Eq.~(\ref{eq:BCRE-h(a)}).
In sectors (I) and (III) the conditions of
Eq.~(\ref{eq:shock-env-def}) cannot be satisfied and thus the
characteristics in these sectors never form an envelop.  This can be
seen as follows. In sector (I), the characteristics along which
$\alpha$ is constant are given by
\begin{equation}
  \label{eq:shock-cc-I}
  x_\alpha(t)=\alpha+R_0(\alpha) \frac{e^{(m + 1/\delta)t}-1}
{m + 1/\delta} \, .
\end{equation}
Using this expression, the second condition of
Eq.~(\ref{eq:shock-env-def}) becomes
\begin{equation}
  \label{eq:shock-cond-I}
  0=1+ \frac{R'_0(\alpha)}{m+1/\delta}
\left[e^{(m+1/\delta)t} - 1 \right]\, .
\end{equation}
For $m+1/\delta > 0$, the term added to one on the rhs is positive for
$t>0$ since $R'_0(\alpha)>0$ for negative $\alpha$ in sector (I).  In
the opposite case $m+1/\delta < 0$, the same argument applies since
the expression in the square brackets in now negative for $t>0$.  This
shows that the rhs is always larger than one, and the condition is
never fulfilled. In sector (III) we use a different argument to show
that no $\alpha$=constant characteristics, which originate from
positive $x$ at $t=0$, form an envelope. From the characteristics in
sector (III), the conditions of Eq.~(\ref{eq:shock-env-def}) are
formally fulfilled by the expression
\begin{subequations}
  \label{eq:shock-env-III}
  \begin{eqnarray}
  t_e(\alpha)&=&\frac{1}{m - 1/\delta}
\ln\left[ 1 - \frac{m - 1/\delta}{R'_0(\alpha)} \right]\\
x_e(\alpha)&=&\alpha-\frac{R_0(\alpha)}{R'_0(\alpha)}
\end{eqnarray}
\end{subequations}
for the envelope. However, it remains to be checked that this curve is
indeed located in sector (III), i.e., if its coordinates are larger
than the boundary between sectors (II) and (III),
cf.~Eq.~(\ref{eq:bound-II-III}), which yields
\begin{equation}
  \label{eq:shock-cond-III}
  x_e(\alpha) > \frac{r_0}{m-1/\delta}\left[ e^{(m-1/\delta)
t_e(\alpha)} -1 \right].
\end{equation}
Using the definition of $R_0(x)$, cf.~Eq.~(\ref{eq:BCRE-R0}), and
the relation $R'_0(\alpha)= -(1/\delta) R_0(\alpha)/(1+ R_0(\alpha))$,
the latter condition turns out to be equivalent to
\begin{equation}
  \label{eq:shock-cond-III-b}
  \ln\left( \frac{R_0(\alpha)}{r_0}\right)+ \frac{r_0}{R_0(\alpha)} < 1.
\end{equation}
Since $R_0(\alpha)/r_0 < 1$ for $\alpha \neq 0$, this condition is in fact
never fulfilled which proves the absence of shocks in sector (III).

Knowing that shocks, i.e., the cusp of the envelope, can occur only in
sector (II) we can try to obtain the condition for shock generation
and the time and position of the shock. First consider a negative
slope $m<0$. Then the characteristics in sector (II) saturate at large
times,
\begin{equation}
  \label{eq:shock-sat-II}
  \lim_{t \to \infty} x_\alpha(t) = \frac{r_0}{1/\delta-m}\left[
1-\frac{m}{r_0} \alpha -\frac{1}{r_0}\ln\left(\frac{h(\alpha)}
{r_0} \right)\right].
\end{equation}
Since $h(\alpha)$ is a monotonously decreasing function for negative
$\alpha$, the expression in the square brackets is monotonously
increasing in $\alpha$. Thus the characteristics retain the original
order for all times, i.e., they never intersect. The situation is more
complicated for positive $m$. Let us assume that there exists a finite
value $m_c$ so that only for $m>m_c$ shocks are produced. Then, one
expects that at $m=m_c$ the time $t_s$ for the shock appearance tends
to infinity in order to have no shocks at finite times for $m<m_c$.
Since the shock position $(t_s,x_s)$ is the cusp of the envelope, we
have to analyze the large time behavior of the envelope of
Eq.~(\ref{eq:env-II}). We start with the assumption that the critical
value $m_c < 1/\delta$, and, in fact, at the end we will confirm this
assumption. For $m<1/\delta$ one has the asymptotic behavior
$t_e(\alpha) \sim -\alpha$, and thus we consider large negative values
for $\alpha$ in the following. The shock time is given by the minimal
time of the envelope $t_s=t_e(\alpha_m)$ with $\alpha_m$ the parameter
at the minimum, i.e., $d t_e(\alpha)/d\alpha =0$ for
$\alpha=\alpha_m$. Then close to the critical slope $m_c$, we expect
that $\alpha_m \to -\infty$. For large negative $\alpha$, the function
$h(\alpha)$ of Eq.~(\ref{eq:BCRE-h(a)}) has the asymptotic form
\begin{equation}
  \label{eq:shock-h-asympt}
  h(\alpha)\simeq r_0^{1-\nu} (-m\alpha)^\nu e^{r_0(1-\nu)(1+\alpha/\delta)}
\end{equation}
with $\nu=(2/\delta)/(m+1/\delta)$.  Using this expansion in
Eq.~(\ref{eq:env-II}) the condition $d t_e(\alpha)/d\alpha =0$ becomes
at asymptotically large $\alpha$ independent of $\alpha$ and assumes
the simple form
\begin{equation}
  \label{eq:shock-cond-asympt}
   m = \frac{\nu-1}{\delta} \, .
\end{equation}
Since we assumed that the shock time $t_s \to \infty$, this condition
has to be regarded as an equation for the critical slope $m=m_c$.
Since $\nu$ depends on $m$ the equation is quadratic in $m$, and it
has one negative solution which we have to discard and the other
solution gives the critical slope beyond which shocks occur,
\begin{equation}
  \label{eq:shock-cond}
  m_c=\frac{\sqrt{2}-1}{\delta} \, .
\end{equation}
This is the result given in Eq.~(\ref{eq:BCRE-shock_m}). The behavior
of $t_s$ close to $m=m_c$ can be obtained by computing the leading
correction to the (constant) asymptotic expression for
$dt_e(\alpha)/d\alpha$. We find a correction $\sim 1/\alpha$ which in
turn yields the leading order of $\alpha_m$ close to $m_c$,
\begin{equation}
  \label{eq:shock-alpha_m}
  \alpha_m = - \frac{1}{\sqrt{2}} \, \frac{1}{m-m_c} \, .
\end{equation}
Since at large negative $\alpha$ the time coordinate of the envelope
behaves as $t_e(\alpha) \sim -\alpha$ and $t_s=t_e(\alpha_m)$, we obtain
the following power law for the shock time close to criticality,
\begin{equation}
  \label{eq:shock-t_s}
  t_s \sim \frac{1}{m-m_c} \, ,
\end{equation}
as given by Eq.~(\ref{eq:BCRE-shock_sing}). The precise time $t_s$ and
position $x_s$ of the shock is given for $m$ sufficiently close to
$m_c$ by the envelope of Eq.~(\ref{eq:env-II}) at $\alpha=\alpha_m$ of
Eq.~(\ref{eq:shock-alpha_m}). At larger $m$ the coordinates
$(t_s,x_s)$ are difficult to compute. However, at sufficiently large
$m$ closed formulas can be obtained. The reason for this is that for
$m$ larger than some threshold the minimum of $t_e(\alpha)$ is always
at $\alpha_m=0$, i.e., the shock is located on the boundary between
sectors (II) and (III). To see this, we expand the envelope of
Eq.~(\ref{eq:env-II}) now around small negative $\alpha$. This gives
\begin{subequations}
  \label{eq:env-expansion}
  \begin{eqnarray}
  \label{eq:env-expansion-1}
  t_e(\alpha)&=&\frac{1}{m-1/\delta} \ln \left(
\frac{m\delta(1+r_0) - 1}{r_0} \right) +
\frac{1-2 r_0 \delta m}{(1+r_0)(\delta m (1+r_0)-1)}\,
\alpha + {\cal O}(\alpha^2)\\
  \label{eq:env-expansion-2}
x_e(\alpha)&=&r_0 \delta \left(1+\frac{1}{r_0}\right) +
\frac{1-2 r_0 \delta m}{1+r_0} \, \alpha + {\cal O}(\alpha^2) \, .
\end{eqnarray}
\end{subequations}
The minimum of $t_e(\alpha)$ is at $\alpha=0$ if the coefficient of
$\alpha$ in Eq.~(\ref{eq:env-expansion-1}) is negative. The
denominator of this coefficient has to be positive since otherwise the
argument of the logarithm in Eq.~(\ref{eq:env-expansion-1}) would be
negative. Thus if the slope $m$ fulfills the two conditions $m >
1/(2r_0\delta)$ and $m>1/(\delta(1+r_0))$ simultaneously then the
shock position is given by $(t_s,x_s)=(t_e(\alpha=0),x_e(\alpha=0))$.
As mentioned already in Sec.~\ref{sec:constant_b_zero} for $r_0<1$ the
first condition is relevant whereas for $r_0>1$ the latter condition
dominates. Now one may ask if it is possible that $m_c$ is larger than
latter thresholds so that the shock would occur no longer at
$\alpha_m$ of Eq.~(\ref{eq:shock-alpha_m}) but at $\alpha_m=0$, i.e.,
on the boundary between sectors (II) and (III) at rather small times.
In fact, for $r_0<1$ the condition $m_c > 1/(2r_0\delta)$ is never
fulfilled so that the minimum remains at $\alpha_m$ of
Eq.~(\ref{eq:shock-alpha_m}). For $r_0>1$ the condition $m_c
>1/(\delta(1+r_0))$ leads to $r_0 > \sqrt{2}$. In the latter case the
shock occurs always at $\alpha_m=0$ and the new critical slope is given by
$1/(\delta(1+r_0))$.  However, it should kept in mind that the width
of the perturbation $R_0(x)$ of Eq. (\ref{eq:BCRE-R0}) is proportional
to $\delta$ only for $r_0 \lesssim 1$. For larger $r_0 \gtrsim 1$ the
width is proportional to $r_0 \delta$. From the first term
$x_s=x_e(\alpha=0)$ in Eq.(\ref{eq:env-expansion-2}) we thus conclude
that the shock occurs at the uphill end of the avalanche with the
shock position approximately given by the uphill end of the
perturbation at $t=0$.

Finally, we study the generation of shocks for model ${\cal L}$. We do
this by using an approach which is more adapted to the special
structure of the solution of this model. We do not use directly the
definition of Eq.~(\ref{eq:shock-env-def}) but look for a
discontinuity in the profile $R(t,x)$ as a function of $x$. If there
is a jump in $R(t,x)$ at some position $x$ then a shock is generated
and the earliest time where this happens if the shock time $t_s$. We
start from Eq.~(\ref{eq:lin_b_0_t-R-rel}) which gives
\begin{equation}
  \label{eq:shock-L-start}
  R(\alpha,\beta) = R_0(\alpha) + mt \, .
\end{equation}
By taking the derivative with respect to $x$, one obtains 
\begin{equation}
  \label{eq:shock-L-dR-dx}
  \dr_x R = R_0'(\alpha) \, \dr_x \alpha \, .
\end{equation}
Since $R'_0(x)$ remains finite, we have to search for a divergence in
$\dr_x \alpha$. The characteristic parameter $\alpha(t,x)$ can be
obtained from Eqs.~(\ref{RofaabbsmallRb=0}) and
(\ref{eq:lin_b_0_t-R-rel}) which yield
\begin{equation}
  \label{eq:shock-L-alpha-result}
  \alpha(t,x) = x +\frac{m}{2} t^2 + R_0(\alpha) \, t \, ,
\end{equation}
which leads to 
\begin{equation}
  \label{eq:shock-L-alpha-deri}
  \dr_x \alpha = \frac{1}{1-R_0'(\alpha) t} \, .
\end{equation}
Since there is always an interval of values for $\alpha$ where a
localized $R_0(x)$ has a positive derivative, shocks are generated
{\it always}. The time scale $t_s$ for the occurrence of the shock is
the earliest time where $\dr_x\alpha$ diverges, i.e., it is given by
the maximal slope of the initial perturbation,
\begin{equation}
  \label{eq:shock-L-ts}
  t_s = \frac{1}{{\rm max}\, R_0'(x)} \, ,
\end{equation}
which is Eq.~(\ref{eq:lin_profile_b_zero_shock_time}). The position
$x_s$ of the shock follows from Eq.~(\ref{eq:shock-L-alpha-result}).
For the Gaussian perturbation $R_0(x)=r_0 \exp(-x^2/\delta^2)$
discussed in Section \ref{sec:linear_b_zero} one has $t_s=\sqrt{e/2}
\, \delta/r_0$ and the position is given by
\begin{equation}
  \label{eq:shock-L-position}
  x_s = -\sqrt{2} \delta - \frac{e}{4} \, 
\left(\frac{\delta}{r_0}\right)^2  \, m \, .
\end{equation}
Note that the shock occurs at the downhill front of the avalanche as
opposed to the uphill position in model ${\cal P}$.

\end{appendix}


\end{document}